\documentclass{emulateapj}

\def \fred{\hbox{$f_{red}$}}
\def \M05{\hbox{$^{0.5}M_{r'}$}}
\def \M05star{\hbox{$^{0.5}M_{r'}^*$}}
\def\gtapr {\lower .1ex\hbox{\rlap{\raise .6ex\hbox{\hskip .3ex
        {\ifmmode{\scriptscriptstyle >}\else
                {$\scriptscriptstyle >$}\fi}}}
        \kern -.4ex{\ifmmode{\scriptscriptstyle \sim}\else
                {$\scriptscriptstyle\sim$}\fi}}}
\def\ltapr {\lower .1ex\hbox{\rlap{\raise .6ex\hbox{\hskip .3ex
        {\ifmmode{\scriptscriptstyle <}\else
                {$\scriptscriptstyle <$}\fi}}}
        \kern -.4ex{\ifmmode{\scriptscriptstyle \sim}\else
                {$\scriptscriptstyle\sim$}\fi}}}

\shortauthors{I.H. Li et al.}
\shorttitle{galaxy evolution at $0.25\leq z \leq 0.75$}

\begin{document}
\title{The WiggleZ Dark Energy Survey: galaxy evolution at $0.25\leq z \leq 0.75$ using the Second Red-Sequence Cluster Survey (RCS-2)}
%
\author{
 I.H. Li\altaffilmark{1,2,*}, H.K.C. Yee\altaffilmark{2},
 Chris Blake\altaffilmark{1}, Sarah Brough\altaffilmark{3},
 Matthew Colless\altaffilmark{3}, Carlos Contreras\altaffilmark{1},
 Warrick J. Couch\altaffilmark{1}, Scott M. Croom$^4$, Tamara
Davis$^{5,6}$, Michael J.\ Drinkwater$^5$, Karl Forster$^7$, David 
G.\ Gilbank$^8$, M.~G.~Gladders$^{9}$, 
Bau-ching Hsieh$^{10}$,
Ben Jelliffe$^{4}$,
Russell J.\ Jurek$^{11}$, 
Karl Glazebrook$^{1}$,
Barry Madore$^{12}$, D.~Christopher  Martin$^7$, Kevin Pimbblet$^{13}$, Gregory B.\ Poole$^1$, Michael
Pracy$^4$, Rob Sharp$^3$, Emily Wisnioski$^1$, David Woods$^{14}$ and Ted
Wyder$^7$} 
\altaffiltext{}{
\noindent
   $^1$ Centre for Astrophysics \& Supercomputing, Swinburne University of Technology, P.O. Box 218, Hawthorn, VIC 3122, Australia \\ 
   $^{2}$ Department of Astronomy and Astrophysics, University of Toronto, 50 St.\ George Street, Toronto, ON M5S 3H4, Canada \\ 
   $^3$ Australian Astronomical Observatory, P.O. Box 296, Epping, NSW 1710, Australia \\
   $^4$ Sydney Institute for Astronomy, School of Physics, University of Sydney, NSW 2006, Australia \\
   $^5$ Department of Physics, University of Queensland, Brisbane, QLD 4072,
   Australia \\ 
   $^6$ Dark Cosmology Centre, Niels Bohr Institute, University of Copenhagen, Juliane Maries Vej 30, DK-2100 Copenhagen, Denmark \\ 
   $^7$ California Institute of Technology, MC 405-47, 1200, East California Boulevard, Pasadena, CA 91125, United States \\ 
   $^8$ Astrophysics and Gravitation Group, Department of Physics and
   Astronomy, University of Waterloo, Waterloo, ON N2L 3G1, Canada \\
   $^{9}$ Department of Astronomy and Astrophysics, University of
   Chicago, 5640 South Ellis Avenue, Chicago, IL 60637, United States \\
   $^{10}$ Institute of Astronomy and Astrophysics, Academia Sinica, PO Box 23-141, Taipei 106, R.O.C. Taiwan \\
   $^{11}$ CSIRO Astronomy \& Space Sciences, Australia Telescope National Facility, Epping, NSW 1710, Australia \\ 
   $^{12}$ Observatories of the Carnegie Institute of Washington, 813 Santa Barbara St., Pasadena, CA 91101, United States \\ 
   $^{13}$ School of Physics, Monash University, Clayton, VIC
   3800, Australia \\ 
   $^{14}$ Department of Physics \& Astronomy, University of British Columbia, 6224 Agricultural Road, Vancouver, BC V6T 1Z1, Canada \\ 
   }

\email{*email:tli@astro.swin.edu.au}


\begin{abstract}
We study the evolution of galaxy populations around the spectroscopic WiggleZ sample of star-forming galaxies at $0.25 \leq z \leq 0.75$ using the photometric catalog from the Second Red-Sequence Cluster Survey (RCS2).
We probe the optical photometric properties of the net excess neighbor galaxies.
The key concept is that the marker galaxies and their neighbors are located at the same redshift, providing a sample of galaxies representing
a complete census of galaxies in the neighborhood of star-forming galaxies. 
The results are compared with those using the RCS WiggleZ Spare-Fibre (RCS-WSF)
sample as markers, representing galaxies in cluster environments
at $0.25\leq z \leq 0.45$. 
By analyzing the stacked color-color properties of the WiggleZ neighbor galaxies, we find that their optical colors are not a strong function of indicators of star-forming activities such as EW([OII]) or GALEX $NUV$ luminoisty of the markers.
The galaxies around the WiggleZ markers exhibit a bimodal distribution on the color-magnitude diagram, with most of them located in the blue cloud.
The optical galaxy luminosity functions (GLF) of the blue neighbor galaxies have a faint-end slope $\alpha$ of $\sim-1.3$, similar to that for galaxies in cluster environments drawn from the RCS-WSF sample.
The faint-end slope of the GLF for the red neighbors, however, is $\sim-0.4$, significantly shallower than the $\sim-0.7$ found for those in cluster environments.
This suggests that the build-up of the faint-end of the red sequence in cluster environments is in a significantly more advanced stage than that in the star-forming and lower galaxy density WiggleZ neighborhoods.
We find that the 
red galaxy fraction ($f_{red}$) around the star-forming WiggleZ galaxies has similar values from  $z\sim0.3$ to $z\sim0.6$ with  $f_{red}\sim0.28$, but drops to $f_{red}\sim0.20$ at $z\gtapr0.7$. 
This change of \fred~with redshift suggests that there is either a 
higher rate of star-forming galaxies entering the luminosity-limited 
sample at $z\gtapr0.7$, or a decrease in the quenching rate of star formation
at that redshift.
Comparing to that in dense cluster environment, the $f_{red}$ of the WiggleZ neighbors is both considerably smaller and has a more moderate change with redshift, pointing to the stronger and more prevalent environmental influences on galaxy evolution in high-density regions.
\end{abstract}

\section{Introduction}
The subject of galaxy evolution has been widely studied using both photometric and spectroscopic data  over wide redshift ranges.
In the Northern sky, the SDSS survey has mapped out a vast region of the
nearby Universe, and numerous studies have investigated galaxy properties 
and environmental influences using its data.
Other surveys, such as COMBO-17 \citep{2003A&A...401...73W} and COSMOS \citep{2007ApJS..172....1S, 2007ApJS..172...70L}, have spent much effort to 
explore galaxy properties and evolution in the more distant universe,
out to redshift $\sim$1 and beyond.
It has become clear that galaxy colors exhibit a bimodal distribution at all redshifts to at least $z\sim1$, with a relatively narrow red sequence dominated by non-star-forming galaxies and a blue cloud of star-forming galaxies \citep[e.g.,][]{2001AJ....122.1861S, 2003ApJ...594..186B, 2004ApJ...608..752B, 2006ApJ...647..853W}.
The fraction of red-sequence galaxies (or blue cloud galaxies) changes 
in different environments and at different redshifts.
It has been found that red passive galaxies tend to populate dense environments and blue star-forming galaxies are more common in less dense regions \citep[e.g.,][]{1980ApJ...236..351D, 2007MNRAS.376.1445C,2009ApJ...698...83L}.
In clusters from $z<0.1$ to $z\sim$0.5, the fraction of blue galaxies increases from a few per cent to $\sim$30\%, and reaches $\sim$70\% at $z\sim$1 \citep[e.g.,][]{1984ApJ...285..426B, 2008ApJ...680..214L, 2009MNRAS.400..687M, 2009ApJ...704..126H}.

It is believed that the red sequence in galaxy clusters is assembled from the top down, being already largely in place at the bright end by $z\sim1$, with 
the faint-end filled in at a later time \citep[e.g.,][]{2004ApJ...608..752B,2005MNRAS.362..268T, 2006ApJ...647..853W, 2007ApJ...661...95S, 2008ApJ...673..742G, 2009MNRAS.400...68D}.
Since there are relatively fewer stars formed in red-sequence galaxies, 
the build-up of the red sequence has been argued to be driven by the global suppression of star formation through environmental-related processes, 
such as galaxy merging, galaxy harassment, gas stripping,
or gas consumption by star-forming disks \citep[e.g.,][]{1983ApJ...270....7D,1991ApJ...370L..65B, 1996Natur.379..613M}.
The build-up of the red sequences in galaxy clusters can be considered as the gradual loss of late-type progenitors over a Hubble time \citep[e.g.,][]{2001ApJ...553...90V, 2005MNRAS.360...60K}.

In field environments, galaxies also exhibit a bimodality in their color distributions, and form a red sequence.
However, the fraction of field galaxies on the red sequence is greatly lower than that in clusters. 
Nevertheless, there is also a deficit of faint red-sequence field galaxies, both at higher redshift \citep[e.g.,at $z\sim$0.8,][]{2005MNRAS.362..268T, 2009A&A...507..671T,2005ApJ...620..595W}
and at the present day \citep[e.g.,][]{2007ApJS..173..293W,2006ApJ...648..268B},
indicating that the assembly of the red sequence is still incomplete in low-density environments.  
As the majority of star formation at all redshifts is contributed by blue late-type galaxies, 
it is interesting to probe galaxy evolution from the perspective of blue star-forming galaxies. 
Especially the average star-formation density is evolving 
rapidly with redshift in the field at least a factor of $\sim$10 since $z\sim$1 \citep[e.g.,][]{1998ApJ...498..106M,2006ApJ...651..142H,2010MNRAS.tmp..704G}.

Many studies of galaxy evolution beyond the local Universe focus on red galaxies or the cluster environment \citep[e.g.,][]{1997ApJ...488L..75B,1998ApJ...501..571G,1998A&A...334...99K,2010ApJ...716..970L}.
While the evidence is clearer on the question of cluster environmental influences, the star formation of galaxies in field regions is still ambiguous.
Because star formation is still active in blue galaxies, 
they provide a more direct observation on the actual dependence of star formation rate on environment and also its evolution.

In this paper we will use a combination of the Second Red-Sequence Cluster
Survey (RCS2; \citep{2007ASPC..379..103Y,2011AJ....141...94G}) 
and the WiggleZ spectroscopic survey \citep{2010MNRAS.401.1429D} to study galaxy evolution up to $z\sim0.7$.
This combination produces one of the largest photometric and spectroscopic databases at intermediate redshifts,
covering a total of $\sim$300 square degrees with $g'r'z'$ photometry
 and optical spectra (4700\AA--9500\AA) for $\sim$120,000 blue star-forming 
galaxies at $0.2 \ltapr z \ltapr 1$.
Such a combined data set provides a great opportunity to explore properties of the galaxy population and its evolution at the intermediate redshift.
The WiggleZ spectroscopic survey targets primarily blue star-forming galaxies,
using UV fluxes as its main selection criteria, along with a set of complex
optical selection rules (see \S\ref{sec:data}).
While the WiggleZ spectroscopic catalog provides a large sample of star-forming
galaxies covering a significant redshift range, its complex 
optical selection criteria make its direct application for investigating
 the evolution of star-forming galaxies complicated, if not impossible.
However, they provide a valuable database as a catalog of {\it markers} of 
regions where star formation is prevalent, which are likely low galaxy 
density regions of the universe.
Inspired by the work of \citet{1987ApJ...319...28Y}, 
who used low-redshift quasars as markers to derive statistically the luminosity function of galaxies associated with quasars,  we approach the task by exploring photometric 
properties of the galaxies around WiggleZ galaxies, which provide an 
unbiased census of galaxies in regions of strong star formation.
The RCS2 survey provides complete and relatively deep optical photometric 
catalogs for half of the fields used by the WiggleZ survey, and they 
are used for the analyses of the WiggleZ galaxy neighbors.
Our work shows that probing the properties of the neighbor galaxies 
of markers statistically can 
offer a powerful method in studying galaxy evolution.

The structure of the paper is as follows.
We describe briefly the WiggleZ spectroscopic project and the photometric RCS2 survey in \S\ref{sec:data}.
Our method of constructing color-color-magnitude cubes of the galaxies associated with the WiggleZ galaxies is detailed in \S\ref{sec:method}.
The results are presented in \S\ref{sec:results}, where we probe the color-color plots, color-magnitude diagrams, luminosity function,  and red-galaxy fractions of the neighbor galaxies.
We discuss the results in \S\ref{sec:discussion} and summarize our work in \S\ref{sec:summary}.
We adopt a cosmology of $\Omega_m$=0.3, $\Omega_{\Lambda}$=0.7, and $H_0$=70km/s/Mpc.

\section{The Surveys and Data} \label{sec:data}

The basic assumption used in this work is that, because galaxies cluster,
 excess galaxies counted around a marker of known redshift are in the same
redshift space as the marker, allowing us to measure their intrinsic photometric
properties such as luminosity and rest-frame colors.
To this end, we require a sample of galaxies with spectroscopic redshifts and
photometric data of the complete field of the spectroscopic sample.
In this section, we first describe briefly the WiggleZ and RCS2 surveys, which
provide the spectroscopic and photometric data, respectively; we then 
present the actual sample of the WiggleZ markers used.
We also present a comparison sample of markers, obtained as part of the
WiggleZ observing runs, based on positions of RCS2 clusters.

\subsection{The WiggleZ Spectroscopic Survey}
\subsubsection{Target Selection}
The WiggleZ Dark Energy Survey is a spectroscopic survey of 240,000 UV-selected emission-line galaxies, designed to map a cosmic volume of $\sim$1Gpc$^3$.
Its primary goal is to precisely measure the scale of baryon acoustic oscillation (BAO) imprinted on the spatial distribution of these galaxies at $0.2 \leq z \leq 1$ \citep[e.g.,][]{2011MNRAS.415.2892B}.
The details of the survey are presented in \citet{2010MNRAS.401.1429D};
here, we present a brief summary.

The survey selects targets from areas totaling $\sim$1,000 deg$^2$ from 7 equatorial regions.
Target galaxies are selected using the $FUV$ and $NUV$ data from the
GALEX Medium Imaging Survey \citep[MIS;][]{2005ApJ...619L...1M}
using the criteria of $FUV-NUV \geq 1$ or no $FUV$ detection.
The targets must also satisfy $NUV \leq 22.8$ and the $NUV$ signal-to-noise
ratio $S/N\geq 3$.
Further selection criteria  based on optical photometry are also applied to 
attempt to
maximize the probability that the targets are blue star-forming galaxies 
at $z\geq0.5$, the primary sample for the WiggleZ project science.
The optical data are obtained from SDSS DR4 \citep{2006ApJS..162...38A} and 
RCS2 (see \S\ref{sec:rcs2}).
In order to select blue star-forming galaxies and exclude spurious matches 
between GALEX and optical data, all WiggleZ targets must also have
$-0.5\leq NUV-r' \leq 2$.
To avoid low-z galaxies, a $20\leq r' \leq 22.5$ criterion together with two different sets of optical color-color selections are applied. 
All these selection criteria give a target density of $\sim$350 galaxies/deg$^2$, or, $\sim$2.6$\pm$0.2\% of optically detected galaxies.
There is no further morphology selection to remove any `stellar' objects, since galactic stars ought to fail the survey selection criteria.

The SDSS data have a depth of [22.0, 22.2, 22.2, 21.3, 20.5] in the $u'g'r'i'z'$ passbands,
while the RCS2 data have a much deeper depth, with average 5$\sigma$ point 
source limits of [24.4, 24.3, 22.8] in the $g'r'z'$ passbands.
Since we want to study the galaxy population properties and their evolution
to as high a redshift as possible, in this paper we use only the RCS2 
regions of the WiggleZ survey.

\subsubsection{Observation and Data}
The WiggleZ observations were conducted using the AAOmega spectrograph
 \citep[the former $2dF$ upgraded;][]{2006SPIE.6269E..14S} on the 3.9m Anglo-Australian Telescope (AAT) from Aug 2006 to Jan 2011.
AAOmega is a fiber spectrograph containing 400 fibers including 8 guide fibers.
Each fiber has a diameter of 2\arcsec.
The field of view is 2 degrees in diameter.
The typical exposure time is 60 min per AAOmega configuration.
This exposure time is too short for allowing a significant detection of galaxy 
continuum for the fainter galaxies, but sufficient to detect emission 
lines for redshift measurement.
Using the 580V and 385R gratings for the blue and red arms with the 670nm dichroic, 
the spectra have a wavelength range from 4700\AA~to 9500~\AA, with a dispersion of $\sim$1.1\AA/pix in the blue arm and $\sim$1.6\AA/pix in the red arm,
providing spectral resolutions of $\sim3.5$\AA~and $\sim5.3$\AA, respectively. 
The observing conditions varied significantly, with the seeing typically ranging from 1-2.5\arcsec.

Detailed descriptions of the data reduction technique and reliability are given
in \citet{2010MNRAS.401.1429D} and summarized here.
The data were reduced during each observing run using the automated $2dFdr$ software developed at the Australian Astronomical Observatory, 
including bias subtraction, flat field, and wavelength calibration.
The redshift of each spectrum was measured using an evolved version of $runz$, which was the software used for 2dFGRS \citep{2001MNRAS.328.1039C} and 2SLAQ \citep{2006MNRAS.372..425C}.
The software has been modified to optimize the use of emission lines to derive redshifts.
The commonly detected emission lines in WiggleZ spectra are [OII]$\lambda$3727,
$H_{\beta}$, [OIII]$\lambda$4959/5007, $H_{\alpha}$, and [NII]$\lambda$6583.
Even though $runz$ automatically generates an integer quality flag ($Q_{zspec}$) in the range of 1-5 based on how well the template fits a given spectrum, 
all WiggleZ spectra were extensively checked visually, and each spectrum was manually assigned a new quality flag.
The redshift confidence increases with larger $Q_{zspec}$.
The redshift reliability has been cross-checked internally using repeated galaxies. 
A subset of redshifts were also compared to DEEP2 galaxies. 
While there may be some debates in distinguishing $Q_{zspec}$=4 and $Q_{zspec}$=5 objects, 
as they are close to being 100\%~reliable,
the critical separation occurs between $Q_{zspec}$=2 and $Q_{zspec}=3$.
For objects with $Q_{zspec}$=3, the redshift reliability is $\sim$79\%.

As of Oct 2009, the survey yielded $\sim$260,000 spectra in total from all 7 equatorial regions, and $\sim$160,000 of them are useful with $Q_{zspec}$=3,4,5.
Part of the WiggleZ spectral database has been released to the public at http://wigglez.swin.edu.au/ds. More details about the project and data can be found in \citet{2010MNRAS.401.1429D}.

\subsection{RCS2 Photometric and Random Catalogs \label{sec:rcs2}}

The RCS2 is a $\sim1000$-square-degree imaging survey in $z'$, $r'$, and
$g'$ with the goal of identifying a large sample ($>10^4$) of galaxy clusters
up to $z\sim1$ for the purpose of constraining cosmological parameters using
the galaxy cluster mass function and studying galaxy evolution.
The survey was carried out using the one-square-degree camera
MegaCam at CFHT.
The survey targets 12 regions of sky with areas varying between 36
to 100 square degrees.
About half of the WiggleZ fields use RCS2 positions and photometry
for target selection.
The three-color photometric catalogs of galaxies in these
targeted fields are used for our analysis of companions
associated with WiggleZ galaxies.
The details for the survey and photometric catalog production
are described in Gilbank et al. (2011); here, we provide a very
brief summary.

The RCS2 photometric catalogs are created using an automated pipeline, 
with algorithms for object finding, photometry, and star-galaxy classification
based on those from the program {\it Picture Processing Program} (PPP)
of \citet{1991PASP..103..396Y}.
The high-precision photometric catalogs in $g'r'z'$ are calibrated using
the colors of the stellar locus combined with overlapping Two-Micron-All-Sky
Survey (2MASS) photometry.
This technique yields an absolute accuracy of better than $\sim0.03$ mag in 
colors, and $\sim0.05$ mag in the $r'$-band, verified via regions that overlap
with the SDSS.
The survey reaches average 5$\sigma$ point source limiting magnitudes
for $z',r',g'$ of 22.8, 24.3, and 24.4, respectively, approximately
2 magnitudes deeper than the SDSS.
Absolute astrometric calibration is accurate to better than $0.3$\arcsec.

A key feature in using the RCS2 catalogs is the availability of 
random catalogs.
These are random points generated to populate the survey area of each
region with a uniform density of one per 10 square arcsecond.
These random points allow one to map out the areas where there are
no data, including chip gaps, bad columns, bright star halos,
saturated pixels, meteor trails, and other cosmetic defects.
These catalogs are crucial in estimating the area sampled by the
data, as in generating proper background count estimates in the
analysis performed in this paper.

\subsection{The Sample of the Markers} \label{subsec:sample}
Since the RCS2 imaging is much deeper than the SDSS data,
the marker galaxies are chosen from the four WiggleZ-RCS2 regions. 
They are RCS2 0047+00, 0310--14, 2143--00, and 2338--09; 
namely, the $01$hr, $03$hr, $22$hr, $00$hr fields in the WiggleZ survey layout.
The use of the RCS2 regions allows us to measure photometric
properties to a much higher redshift.

\begin{figure}
\includegraphics[scale=0.50]{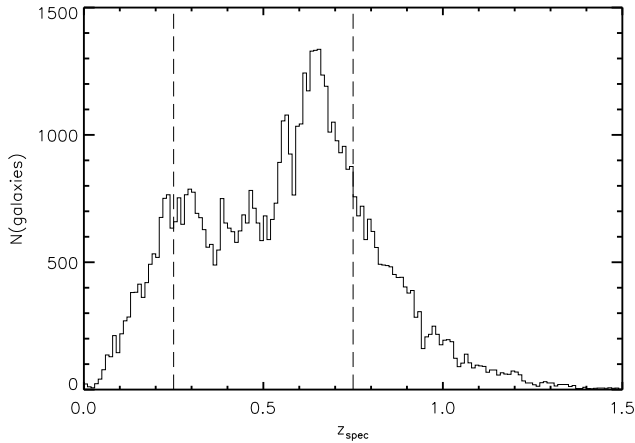}
\caption{ $z_{spec}$ distribution of 62785 WiggleZ galaxies with redshift quality flag $Q_{zspec}$=3,4,5 at $z<1.5$ from four RCS2 patches. The redshift binsize is $\Delta z$=0.01. 
The WiggleZ selection function is designed to optimize observations of high-redshift galaxies thus a significant number of galaxies with $z\sim$0.3--0.6 are removed from the sample.
In this paper we focus on the redshift range between $z$=0.25 and $z$=0.75, which are marked by the vertical dashed lines.
\label{fig:zhis}}
\end{figure}

We use the WiggleZ data taken prior to Oct 2009. 
There are 62785 spectra in total with redshift quality flag $Q_{zspec}\geq 3$ from these four WiggleZ-RCS2 regions.
The redshift distribution is presented in Fig. \ref{fig:zhis}. 
The median redshift is $z\sim$0.59.
We focus on the redshift range between 0.25 and 0.75, giving a total of 41041 spectra.
The lower $z=0.25$ boundary is chosen so that our bluest passband $g'$ is
still blueward of the 4000\AA~break at the lowest redshift bin,
while the upper redshift limit is set based on having sufficient depth
in the imaging data for the analysis of the companion galaxies.

\begin{figure}
\includegraphics[scale=0.37,angle=90]{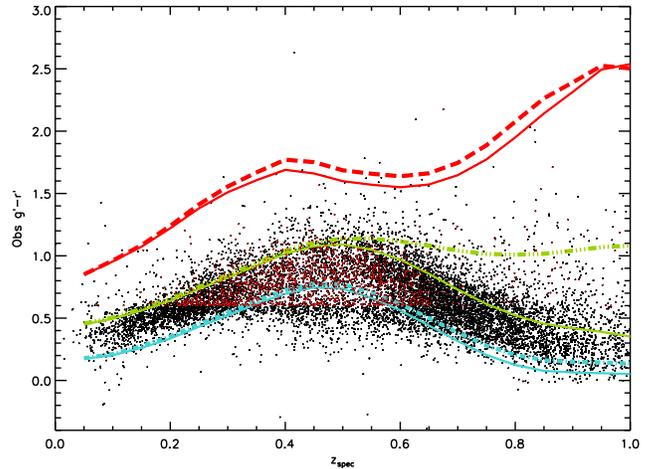}
\caption{The observed $g'$-$r'$ colors as a function of redshift for WiggleZ galaxies. Only 1/5 of the sample is plotted for clarity.
The red, green, and blue solid curves are the colors from GISSEL evolving spetra described in \S2.3, 
while the corresponding (dot-) dashed curves are the colors for the non-evolving spectra used for k-correction.
A `hollow' feature with a deficit of galaxies is observed at $z$=$\sim$0.3--0.6 and $g'$-$r'\sim$0.7 due to the `low-z rejection' in the survey selection criteria. 
The red dots are galaxies that satisfy the `low-z rejection' criteria but were still observed during early stages of the WiggleZ survey (before these criteria were fully implemented).
\label{fig:color_redshift}}
\end{figure}
The observed $g'$-$r'$ colors of the WiggleZ galaxies as a function of redshift are presented in Figure \ref{fig:color_redshift}.
We also overplot three dust-free  models generated from GISSEL \citep{2003MNRAS.344.1000B}. 
The red dashed curve is the Single Stellar Population (SSP) model using the 
Padova \citep{1994A&AS..106..275B} evolutionary tracks with solar 
metallicity $Z$=0.02 and the \citet{2003PASP..115..763C} IMF with a 
zero-redshift age of 13 Gyrs, representing evolved early-type
red galaxies.
The green dot-dashed curve is a $\tau$ model with an exponentially decreasing
star-formation rate with $\tau=1$ Gyr, generated with a metallicity 
of $Z$=0.0001 and a zero-redshift age of 11 Gyr, representing mildly
star-forming spiral galaxies.
The cyan short-dashed line is a model made with a constant star formation rate, representing star-forming late-type/irregular galaxies.

We observe two features from Figure \ref{fig:color_redshift}. 
First, red elliptical galaxies are absent from the WiggleZ sample.
Most of the galaxies populate the region between the $\tau$- and 
constant star-formation models.
This is expected, because the WiggleZ galaxies are selected by the UV flux
to be star-forming galaxies.
 
Second, there is a `hollow' region with a deficit of galaxies at $g'$-$r'$$\sim$0.7 between $z\sim0.3-0.6$. 
This is also manifested as a dip in the redshift distribution of the
WiggleZ markers at $0.3<z<0.6$ in Fig. \ref{fig:zhis}.
This `hollow' feature arises artificially due to the `low-redshift 
rejection' (LRR) criteria based on $g'-r' > 0.6$ and $r'-z' < 0.7(g'-r')$ in
the WiggleZ's  selection criteria used in the RCS2 regions, in the attempt to
maximize $z\geq0.5$ galaxies.
Thus, the LRR criteria actually remove galaxies at redshift up to $z\sim0.6$,
producing the significant broad dip in the redshift distribution of the WiggleZ galaxies.
Nevertheless, the `hollow' region has more galaxies in it than
expected with the LRR criteria.
There are  8597 galaxies in the sample
that actually meet the LRR criteria but are still included in the
WiggleZ sample.
This is because the LRR criteria were developed after the early observing runs 
and thus some galaxies which were initially observed would have been rejected later on by the refined selection criteria.
We find that these galaxies, overplotted in Fig. \ref{fig:color_redshift} as red dots, are primarily distributed over $0.25 \leq z \leq 0.65$.
Thus, while the WiggleZ survey intends to target blue star-forming 
galaxies, an examination of Figure \ref{fig:color_redshift} indicates that the sample is composed of a range of star-forming galaxies, with colors consistent with starburst to constant and mildly star-forming galaxies.

Finally, we note that any direct comparison  of the properties
in this spectroscopic sample as a function of redshift is not 
straightforward, due to the $20\leq r'\leq 22.5$ selection criterion.
This criterion produces galaxy samples of different absolute magnitude ranges
at different redshifts.

\subsection{The RCS2 Cluster WiggleZ Spare-Fibre (RCS-WSF) Sample}
During the WiggleZ observing runs, a small number of AAOmega fibres were used for
targets from different projects, using identical observation parameters and
data reduction techniques.
In addition to the sample of the UV-selected WiggleZ galaxies, there are 
$\sim$3000 spectra targeting RCS2 cluster galaxies as part of 
the WiggleZ-RCS2 collaboration.
These galaxies are selected from a preliminary sample of RCS2 clusters
at $z\leq0.5$, chosen as possible high ranking, bright, red-sequence galaxies 
in the clusters.
Most of these galaxies are at $z<0.45$ and have a mean redshift of $z\sim$0.28. 
The details and scientific results using this WiggleZ-RCS2 cluster subsample will be presented in a future paper. 
For the purpose of this work,
they serve as an excellent comparison sample of markers to the WiggleZ
galaxies, as they are red galaxies in dense environments.
Since these galaxies do not cover the same redshift range as 
the WiggleZ galaxies, the comparison is only available at 
lower redshifts.
We will refer to this sample as the RCS2 WiggleZ Spare-Fibre, or RCS-WSF, sample for the remainder of the paper.
\section{Method} \label{sec:method}
With the assumption that the WiggleZ marker galaxies and their neighbors reside in the same spatial regions,
we can construct a net color-color-magnitude (CCM) cube of the neighbor galaxies, so the colors and magnitude information of the neighbor galaxies can be preserved.
The $xyz$ axes of the cube represent $r'-z'$, $g'-r'$, and $r'$, respectively.
Thus, the net counts as a function of luminosity in the $r'$ passband, for instance, can be computed by summing the values in the $x$ and $y$ axes along the $z$ axis.
Essentially, we adopt the method used in \citet{2008ApJ...673..742G} and \citet{2008ApJ...680..214L}
for creating color-magnitude diagrams, but extend the concept to a 3D cube.
To produce the net CCM cube, we subtract a background CCM cube from the total-count CCM cube.
The cubes are made in both observed and rest frames.
We detail the methods below.

\subsection{Observed Color-Color-Magnitude Cubes}
To construct a CCM cube, we first identify all galaxies in the RCS2 photometric catalogs with $r'\leq 24.0$ within a projected comoving radius of $r_p$ Mpc to a WiggleZ galaxy.
The WiggleZ galaxy itself is excluded in this process.
These galaxies are namely the `neighbors' to the WiggleZ galaxy, and their observed $r'-z'$, $g'-r'$, and $r'$ are gridded into a cube with a binsize of 0.05, 0.05, and 0.1 along the $xyz$ (color-color-magnitude) axes.
The CCM cube of the control field (i.e., the background) is constructed using all galaxies with $r' \leq 24$ in the same RCS2 patch. 
A single  RCS2 patch is sufficiently large (typically
81 square degrees) to provide excellent
background statistics, and by using the same patch, it also
ensures a minimal systematic effect.
The control cube has the same binsize as the cube of the marker neighbors.
The count of each element in this control cube is then scaled by 
$N_{ran,in}/N_{ran,tot}$, which is the ratio of the number of the random points within the aperture ($N_{ran,in}$) to the total count ($N_{ran,tot}$) in the patch (see \S\ref{sec:rcs2}).
The typical $N_{ran,in}/N_{ran,tot}$ is $\sim$ $2\times10^{-6}$.
A net CCM cube is obtained by subtracting the scaled control cube from the cube of the neighbor galaxies, i.e., 
	\begin{displaymath}
	\rm cube_{net} = cube_{neighbor} - cube_{background}\times~N_{ran,in}/N_{ran,tot}. 
	\end{displaymath}
These net CCM cubes from individual markers can then be stacked to form
the total CCM cube.

\subsection{Rest-Frame Color-Color-Magnitude Cubes}
One approach to obtain a rest-frame CCM cube is to convert it from an observed one which has been described above.
However, it requires a large amount of computing time to k-correct each element of an observed CCM cube.
An alternative is to compute the rest-frame magnitude and colors of each galaxy first, then construct the rest-frame CCM cube using the same procedure as building the observed cube.
Since  there is no actual redshift information for the neighbors,
it is  assumed that all surrounding galaxies are at the same redshift as the marker.
A control field cube is computed for each marker, for which we also convert 
all galaxies into ``rest-frame photometry" using the redshift of the WiggleZ galaxy.

\begin{figure}
\includegraphics[scale=0.5]{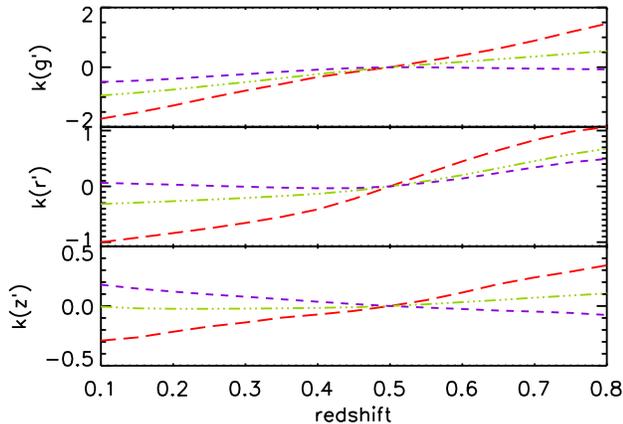}
\caption{k-correction as a function of redshift for $g'r'z'$ filters. The dashed red, dot-dashed green, and short-dashed purple curves are generated using the same model in Fig. \ref{fig:color_redshift}, which represent evolved early-type red, mildly star-forming spiral, and star-forming late-type/irregular galaxies respectively.
\label{fig:kcorrection}}
\end{figure}

The k-correction is derived using tables generated for each of the 
GISSEL \citep{2003MNRAS.344.1000B} models described in \S\ref{subsec:sample}.
Each table contains galaxy colors of the model and 
the k-correction values for each passband as a function of redshift.
For a galaxy at a fixed redshift, we derive the k-correction using the model
grid by interpolating (or extrapolating in some instances) the model colors to match the observed galaxy colors.
The model colors here are the observer-frame colors of GISSEL galaxies with non-evolving spectra, which are overplotted as curves in Figure \ref{fig:color_redshift}.
We use the $g'-r'$ color to derive k-corrections for the $g'$ and $r'$ passbands, and $r'-z'$ for the $z'$ magnitude.
We have compared our k-correction results to the SDSS galaxies in one region, where their k-corrections are available from the official SDSS database. 
Our method in deriving the k-correction yields a good correlation with the SDSS values.
We use $z=0.5$ as our reference redshift, and all the rest-frame photometry is computed relative to this redshift as 
	$^{0.5}M = m - DM - (k-k_{z=0.50})$, 
where $DM$ is the distance modulus.
For reference, Figure \ref{fig:kcorrection} plots the k-correction as a function of redshift for each filter for the three spectral types.

\section{Results} \label{sec:results}
In this section we derive the various photometric properties of the galaxy 
population associated with the WiggleZ marker galaxies.
Our sample is limited to $0.25 \leq z < 0.75$, 
and is divided into five redshift bins with $\Delta z$=0.10.
We refer to these redshift bins as $z=$ 0.3, 0.4, 0.5, 0.6 and 0.7.
We group the WiggleZ galaxies into the redshift bins, and stack all the net CCM cubes
within each redshift bin.
The numbers of the markers and their neighbors in each redshift bin
are listed in Table \ref{tab:count}. 
A total of 45,198 net galaxy counts around 41,041 markers are used in our analysis.
All the CCM cubes are made using galaxies within a projected comoving radius $r_p$=0.25Mpc from the markers. 
The choice of this $r_p$ will be justified in \S4.2.
The CCM cubes of the RCS-WSF sample presented in \S4.5 are made with an angular-diameter radius of $r_p$=0.25Mpc instead of a comoving one, since cluster 
galaxies are considered gravitationally bound, although the results are similar when using either a comoving or angular-diameter radius due to their redshift range.
The galaxy counts in the RCS-WSF sample are tabulated in Table \ref{tab:CLcount}.
The average number of net companions to the markers in this sample are about an
order of magnitude larger than that for the WiggleZ sample.
	\begin{table}
	\caption{Galaxy Counts} \label{tab:count}
	\begin{tabular}{llll}
	\hline\hline
	redshift & $N_{WiggleZ}$ & $N_{net}$ & $N_{background}$ \\
	\hline
	0.25--0.35 & 6885 & 10332.0 & 118828.0 \\
	0.35--0.45 & 6090 & 7811.85 & 61085.1 \\
	0.45--0.55 & 6857 & 8021.05 & 46571.0 \\
	0.55--0.65 & 10869 & 9948.05 & 54039.0 \\
	0.65--0.75 & 10340 & 9084.79 & 40491.2 \\
	\hline\hline
	\footnotetext{Note: $N_{net}$ and $N_{background}$ are within a projected co-moving radius $r_p$=0.25Mpc.}
	\end{tabular}
	\end{table}
	\begin{table}
	\caption{Galaxy Counts for the RCS-WSF Sample} \label{tab:CLcount}
	\begin{tabular}{llll}
	\hline\hline
	redshift & $N_{RCS-WSF}$ & $N_{net}$ & $N_{background}$ \\
	\hline
	0.25--0.35 & 416 & 8968.30 & 12270.7 \\
	0.35--0.45 & 294 & 4760.45 & 5754.54 \\
	\hline\hline
	\footnotetext{Note: $N_{net}$ and $N_{background}$ are within an angular-diameter radius $r_p$=0.25Mpc.}
	\end{tabular}
	\end{table}

\subsection{Random Marker Fields} \label{sec:results-random}
To test the reliability of background subtraction in our method, 
we construct CCM cubes based on the positions of 4000 randomly drawn points from the random catalogs in each of the four RCS2 patch.
These random points are then assigned a redshift between $z$=0.25 and $z$=0.75.
This gives us $\sim$3200 random points in total for each redshift bin.
All the CCM cubes are built in the same way as described in \S\ref{sec:method} but with the marker position and redshift replaced. 
Because these markers are randomly chosen and not based on actual positions 
of any galaxy, we expect the average net excess in the neighbor counts to be zero when these net CCM cubes are stacked, if the background subtraction is properly handled.

\begin{figure}
\includegraphics[scale=0.37,angle=90]{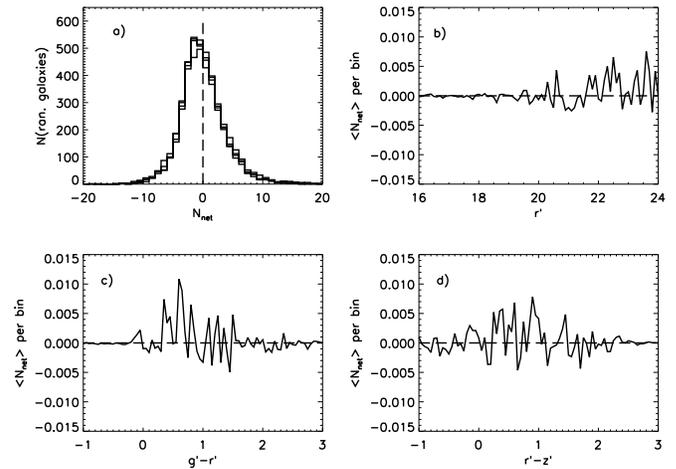}
\caption{The $N_{net}$ of the random catalogs computed using $r_p$=0.25Mpc.
a): distributions of the $N_{net}$ for the 4000 random markers in each RCS2 patch.
b), c), d): The mean $N_{net}$ per bin ($\Delta r'=0.1$, $\Delta(g'-r')=0.05$, or $\Delta(r'-z')=0.05$) along each axis of the stacked cube over $0.25 \leq z < 0.75$. No systematic offset is observed.
\label{fig:ranplot}}
\end{figure}

We compute a net neighbor count, $N_{net}$, for each random point by summing the intensity in all elements of an observed-frame net CCM cube built with $r_p$=0.25Mpc.
The $N_{net}$ distribution together with its dependence on galaxy magnitudes and colors are plotted in Fig. \ref{fig:ranplot}.
Panel a) shows the $N_{net}$ distributions of the 4000 random markers in each RCS2 patch.
These distributions are statistically identical for the different patches.
Summing these distributions gives 
a median  $N_{net}= -0.344$ and a mean $N_{net}= 0.032\pm0.027$ galaxies,
where the uncertainty is the rms of the mean.
Thus, the mean of the net counts around random points is consistent with
being zero.
The relatively large negative value of the median of the net counts is
the result of galaxies being clustered even on the projected sky, 
which results in a skewed histogram of the net count distribution.
Because there is no observed offset in $<N_{net}>$ among different redshift bins, 
we stack all the cubes over $0.25 \leq z < 0.75$, and project the stacked cube along an axis of $r'$, $g'-r'$, or $r'-z'$.
The total $N_{net}$ of the stacked cube along each axis is presented in the b), c), and d) panels in the Figure.
The $<N_{net}>$ counts are  also not a function of magnitude and colors,
and have means of essentially 0, indicating that the background 
contamination is correctly subtracted, statistically speaking.
Given these results, we are confident of our method in background correction and constructing the CCM cubes.

\subsection{Net Excess Galaxy Surface Density} \label{sec:results-density}
The WiggleZ survey targets blue star-forming galaxies with a set of
 complex selection functions.
Blue star-forming galaxies are believed to populate  less dense environment compared to red passive galaxies \citep[e.g.,][]{1980ApJ...236..351D,2006MNRAS.366....2W,2007MNRAS.376.1445C}.
To investigate the characteristics of the neighborhood of WiggleZ galaxies,
we probe the total net neighbor counts, $N_{net}$, as a function of radius centered at each WiggleZ galaxy.
The observed CCM cubes in a series of annuli are computed, and $N_{net}$ in 
an annulus is the sum of the intensity of all elements in the cube. 
Even though all the observed CCM cubes are constructed using galaxies to a fixed
apparent magnitude of $r'$=24.0, we note that the comparison among different annuli at a fixed redshift bin is still meaningful. 
Direct comparisons among different redshift bins, however,
cannot be made because the cubes are not limited to the same 
absolute magnitude depth for the different redshift bins. 
\begin{figure}
\includegraphics[scale=0.37,angle=90]{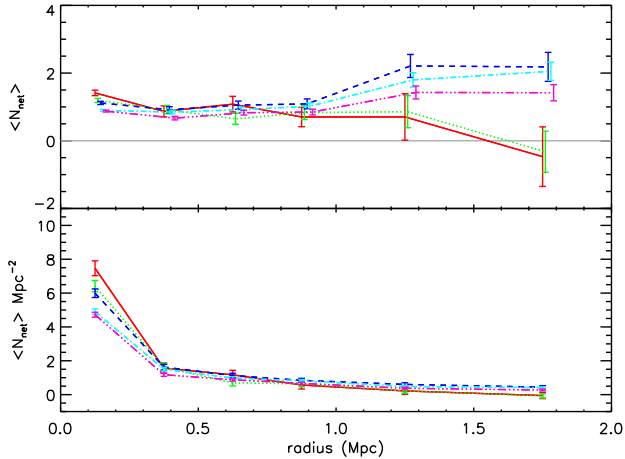}
\caption{The differential mean $N_{net}$ in each annulus without (top) and with (bottom) area normalization. In each panel, the solid red, dotted green, dashed blue, dot-dashed cyan, and dot-dot-dashed pink curves represent our redshift bins in an increasing order.
There are $\sim$1.5 net excess galaxies within $r_p$=0.25Mpc, giving a surface density of $\sim$6 gal/Mpc$^2$.
The surface density drops quickly to $\sim$0 gal/Mpc$^2$ at $r_p\geq1.5$Mpc.
\label{fig:nnetmean}}
\end{figure}
The mean $N_{net}$ in each annulus for the different redshift bins
are plotted in Fig. \ref{fig:nnetmean} as a function of $r_p$.
The number of net excess galaxies within an annulus is not large, the maximum
being only $\sim$1.5 within $r_p$=0.25 Mpc.
Normalizing the $N_{net}$ by the aperture size, 
the mean surface density is a strong function of radius, 
being $\sim$6 gal/Mpc$^2$ within $r_p$= 0.25Mpc and then decreasing rapidly with increasing radius and reaching $\sim$0 at $r_p \geq 1.5$~Mpc. 
Because most $N_{net}$ excess is observed within 0.25 Mpc, we therefore use $r_p$=0.25 Mpc to construct the observed- and rest-frame CCM cubes for our further analysis.

\subsection{The WiggleZ Galaxies and their Neighbors}
\subsubsection{Observed Color-Color Diagrams} \label{sec:obsccm}
Observationally speaking, galaxies appear to be primarily divided into two 
classes. 
One is red passive galaxies and the other, blue star-forming galaxies.
These two classes of galaxies form the so-called `red sequence' and `blue cloud' in color-magnitude space.
In fact, red galaxies may be a mix of truly old passive galaxies and dusty 
star-forming galaxies, and they cannot be distinguished well using a single optical color.
However, \citet{2005A&A...443..435W} showed that dusty star-forming galaxies 
can be well separated from old passive ones in a color-color space, as long 
as one color brackets the 4000\AA~break and the other is at a longer wavelength.
They found that dusty red galaxies actually form a continuous tail extending
from the blue cloud, while old red galaxies form a separate structure of their own (the red sequence).
This makes the color-color diagram a powerful diagnostic tool.

\begin{figure*}
\includegraphics[scale=0.74,angle=90]{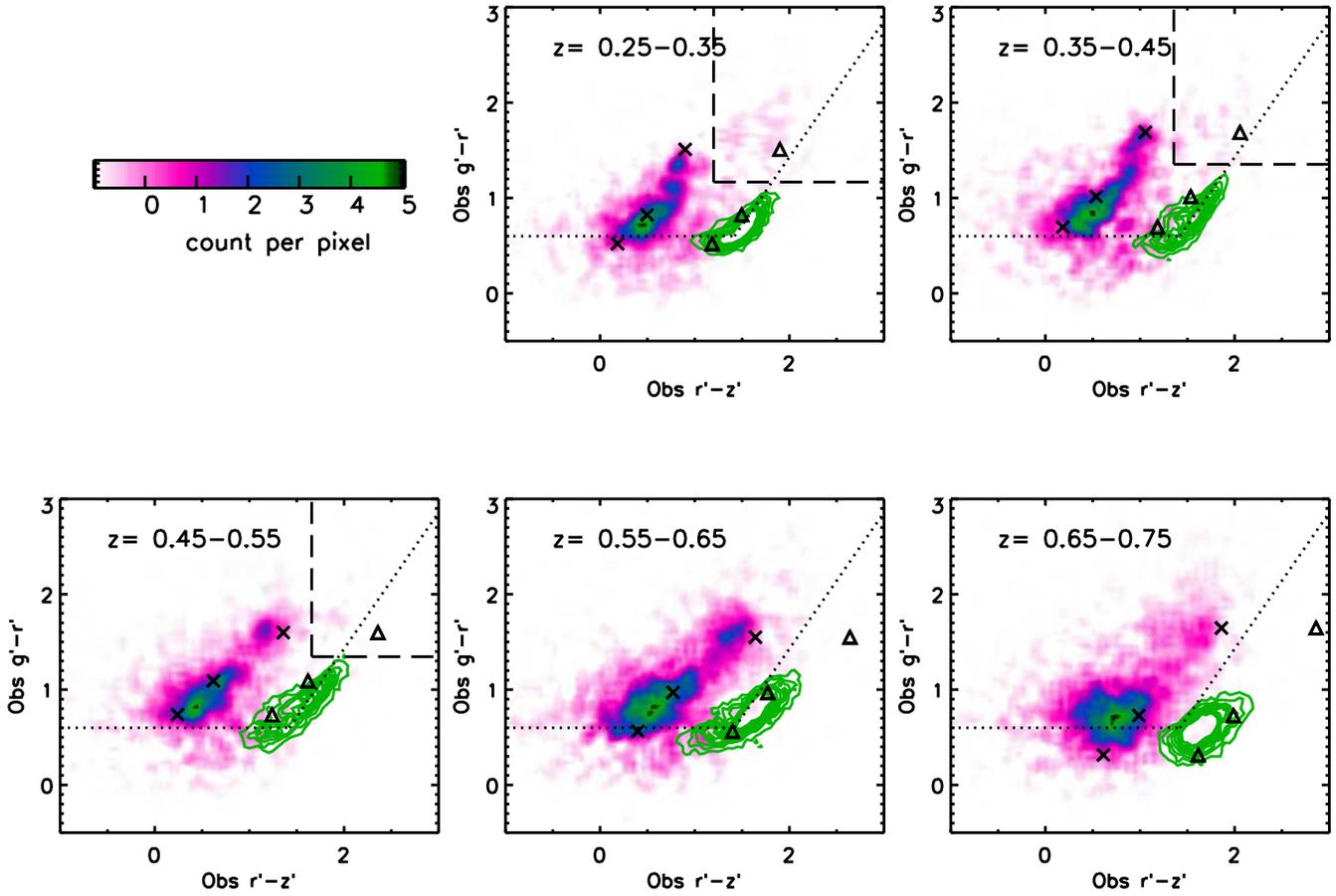}
\caption{Observed color-color diagrams at five redshift bins. The total net count is normalized to 1$\times10^4$ in each panel. The intensity plot is in counts
per pixel of 0.0125$\times$0.0125 mag. The WiggleZ galaxies are plotted as the non-filled contours with a level of 10 galaxies, but shifted by $\Delta$($r'$-$z'$)=1 for clarity. 
The WiggleZ LRR criteria are shown as the black dotted lines, with the $r'-z'$ shifted as well.
The crosses are the colors of three model galaxies with evolving spectra  
shown in Fig. \ref{fig:color_redshift}, and their colors shifted by $\Delta$($r'$-$z'$)=1 are plotted as the triangles.
Both the WiggleZ galaxies and the neighbors form a continuous sequence at each redshift, with the majority of them populated in a blue region. The contamination of dusty star-forming galaxies, indicated by the dashed box, is insignificant for both the markers and the neighbors.
\label{fig:obsccplot}}
\end{figure*}

Figure \ref{fig:obsccplot} presents the observed color-color diagrams for
neighbors in 5 redshift bins, where the colors of the three models 
from Figure \ref{fig:color_redshift} are overlaid as crosses for reference.
For better visual presentation, the pixels
($\Delta(g'-r') \times \Delta(r'-z')$) in the color-color intensity
plot are  subdivided 
by a factor of 4 into smaller pixels in units of 0.0125 mag, and then
smoothed by a kernel of 10$\times$10 pixels. 
The intensity scale is in units of counts per small pixels after
normalizing the net counts to $1\times10^4$ in each redshift bin.
We also overplot the WiggleZ galaxies as the non-filled contours with 
a $\Delta$($r'$-$z'$)=1 offset for clarity.
We observe that, for all redshift bins, both the WiggleZ galaxies and most of their neighbors populate a similar color-color plot, with the exception that
the WiggleZ galaxies do not show a clump of red-passive galaxies.
They both exhibit a continuous sequence in all redshift divisions.
The sequence runs from the blue star-forming regions toward the red passive area, marked by the model colors; 
but no WiggleZ markers have colors as red as the red passive galaxies, indicating that they contain little dust.
We note that this is likely a reflection of the survey design, as the WiggleZ galaxies are selected primarily by UV fluxes.

Although most of the neighbor galaxies reside in the star-forming sequence, some neighbors populate the region of passive red galaxies. 
These red neighbors are red-sequence galaxies, and we will discuss their properties later in the paper.
Some neighbors in the lower-redshift bins, however, exhibit redder ($z'-r'$)
colors than that expected from passive red galaxies at their fixed redshift,
and form roughly a continuous sequence from the blue star-forming galaxies.
These neighbors are possible dusty galaxies. 
We note that such estimation is approximate, since the regions in the 
color-color diagram for dusty reddened star-forming galaxies may change at 
different redshifts due to the shifting of the 4000\AA~break in observed frame.
At $z\sim0.6$ and beyond,
the use of the current color-color diagram to distinguish between dusty 
star-forming and red passive galaxies is not optimal,
because the 4000\AA~break is shifted beyond the center of the $r'$ passband, 
and having only one passband ($z'$) at longer wavelengths is not sufficient 
to distinguish between passive and dusty SEDs.
Using a rough color cutoffs 
to define dusty galaxies as $(r'-z')-C_E>0.3$ and $g'-r' \geq \Delta C$ where $C_E$ is the color of the red elliptical model 
in Figure \ref{fig:color_redshift} and $\Delta C$ is the color halfway 
between the red and green models in the same Figure,
we estimate about 5.2$\pm$1.0\%, 2.4$\pm$0.8\%, and 0.2$\pm$0.5\%~of the
galaxies in the three lower-redshift bins in increasing redshift order 
may be dusty star-forming galaxies.
 From these fractions, we conclude that dust reddened star-forming galaxies 
are not likely a significant component in the WiggleZ neighbors, at least for $z \ltapr 0.6$.
Note that the decrease in the dusty galaxy fraction is likely due to the 
shifting of the $r'$ towards the 4000\AA~break and possibly the different
luminosity depths in the redshift bins, and does not necessarily
reflect a real change.

\subsubsection{Control for the Star-Formation Rate of the Markers} \label{sec:obsccm_line}
Since the star-formation properties of the WiggleZ galaxies themselves
are not uniform across redshift due to the complex selection criteria,
we want to examine whether the properties of the neighbors vary with the 
star-formation activity of the WiggleZ markers.
We use the [OII]$\lambda$3727\AA~emission line equivalent width (EW) of the WiggleZ markers as the proxy for the specific star-formation rate of the markers
and probe where neighbors around markers with different EW([OII]$\lambda$3727\AA) populate the observed color-color diagrams.

First, we measure the [OII]$\lambda$3727\AA~equivalent width of the markers.  
The observed spectra have weak continuum due to the short exposure time ($\sim$1hr), hence the measured EW is noisy for the fainter galaxies.
To measure the EW([OII]$\lambda$3727\AA), we define a window of 10\AA~centered at 3727.8\AA~as the region for the [OII]$\lambda$3727\AA line.
The continuum level is determined using a window of 30\AA~on each side of the [OII] line (on the rest frame), starting at 3677.8\AA~and 3757.8\AA.
Either a linear or second-order polynomial function, whichever returns the smallest $\chi^2$, is chosen to describe the fitted continuum within the windows. 
The spectrum is then subtracted by this fitted continuum. 
A bi-Gaussian function is then applied based on the data points within all three windows. 
The flux of the [OII]$\lambda$3727\AA~line is accordingly the total net flux under this fitted bi-Gaussian curve.
We divide the WiggleZ galaxies into three bins based on the 33.3\% percentiles of the EW([OII]$\lambda$3727\AA) at each redshift bin.
Galaxies without any [OII]$\lambda$3727\AA~detection are excluded.
The median and the 1$\sigma$ uncertainty of the computed 
EW([OII]$\lambda$3727\AA) are about 112.5$\pm$32.6\AA~and 17.7$\pm$8.0\AA~for 
the highest and lowest EW([OII]$\lambda$3727\AA) bins at $z$=0.25-0.35, respectively.

\begin{figure*}
\includegraphics[scale=0.74,angle=90]{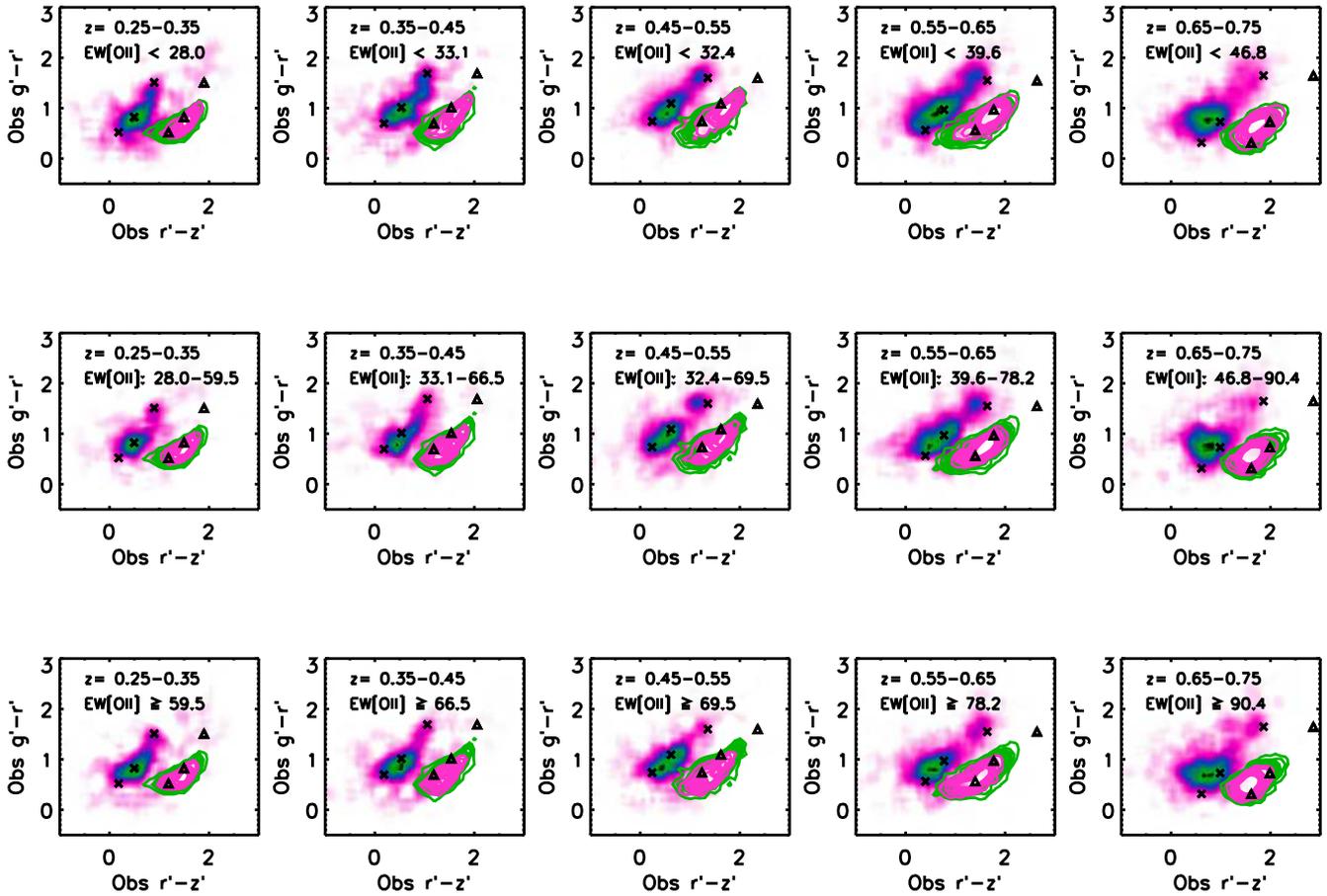}
\caption{Observed color-color diagrams for the WiggleZ and their neighbors where
 the markers are controlled for their EW([OII]$\lambda$3727\AA).
The total intensity for the neighbors in each panel is normalized to a total count of 1$\times10^4$,
 and on the same scale as Fig. \ref{fig:obsccplot}.
The WiggleZ galaxies are plotted as the open pink contours with a level of 10 galaxies. The green contours are those in Fig. \ref{fig:obsccplot} for all the WiggleZ markers at a fixed redshift bin as reference.
Within the same redshift bin, both the $g'-r'$ and $r'-i'$ color distributions
of the WiggleZ galaxies and their neighbors remain similar for
samples with different EW([OII]$\lambda$3727\AA), indicating that
the specific star-formation rate of the parent markers does not affect
the properties of the neighbors.
\label{fig:ccplot_O2}}
\end{figure*}
We re-stack the observed CCM cubes of the neighbors by dividing the sample 
into bins of EW([OII]$\lambda$3727\AA) within each redshift bin
 and present them in 
Figure \ref{fig:ccplot_O2}, where the WiggleZ markers themselves are again offset by $\Delta(r'-z')=1$ for clarity.
 From Figure \ref{fig:ccplot_O2}, it is clear that 
the color-color distributions for neighbors of WiggleZ galaxies of different EW([OII]) are very similar within the same redshift bin.
Kolmogorov-Smirnov tests of pairs of both $g'-r'$ and $r'-z'$
distributions find no significant difference between the different 
EW([OII]$\lambda$3727\AA) bins within the same redshift bin.
The smallest significant level in all the pair-wise comparisons is 0.22.
This implies that the properties of the neighbors do not strongly 
depend on the properties of the WiggleZ galaxies.
Furthermore, we find that the $N_{net}$ distributions of the neighbors within 0.25 Mpc, although not shown here, are identical among all WiggleZ galaxies with different EW([OII]$\lambda$3727\AA) at a fixed redshift bin.
This adds to the growing evidence that 
environment has little influence in the properties of star-forming galaxies
\citep[e.g.,][]{2004ApJ...615L.101B,2005ApJ...629L..77Y,2001ApJ...559..606C,2005AJ....130.1482R,2007ApJS..172..270C,2009MNRAS.398..754B}.
Therefore, we conclude that 
the insignificant dependence between the properties of the neighbors and the markers allows us to explore galaxy evolution using the neighbors, 
even though the WiggleZ galaxies may cover different ranges of properties
 at different redshifts due to the survey selection criteria; for example,
 the $r'$=20-22.5 criterion naturally select more massive galaxies at higher redshifts. 

\begin{figure*}
\includegraphics[scale=0.74, angle=90]{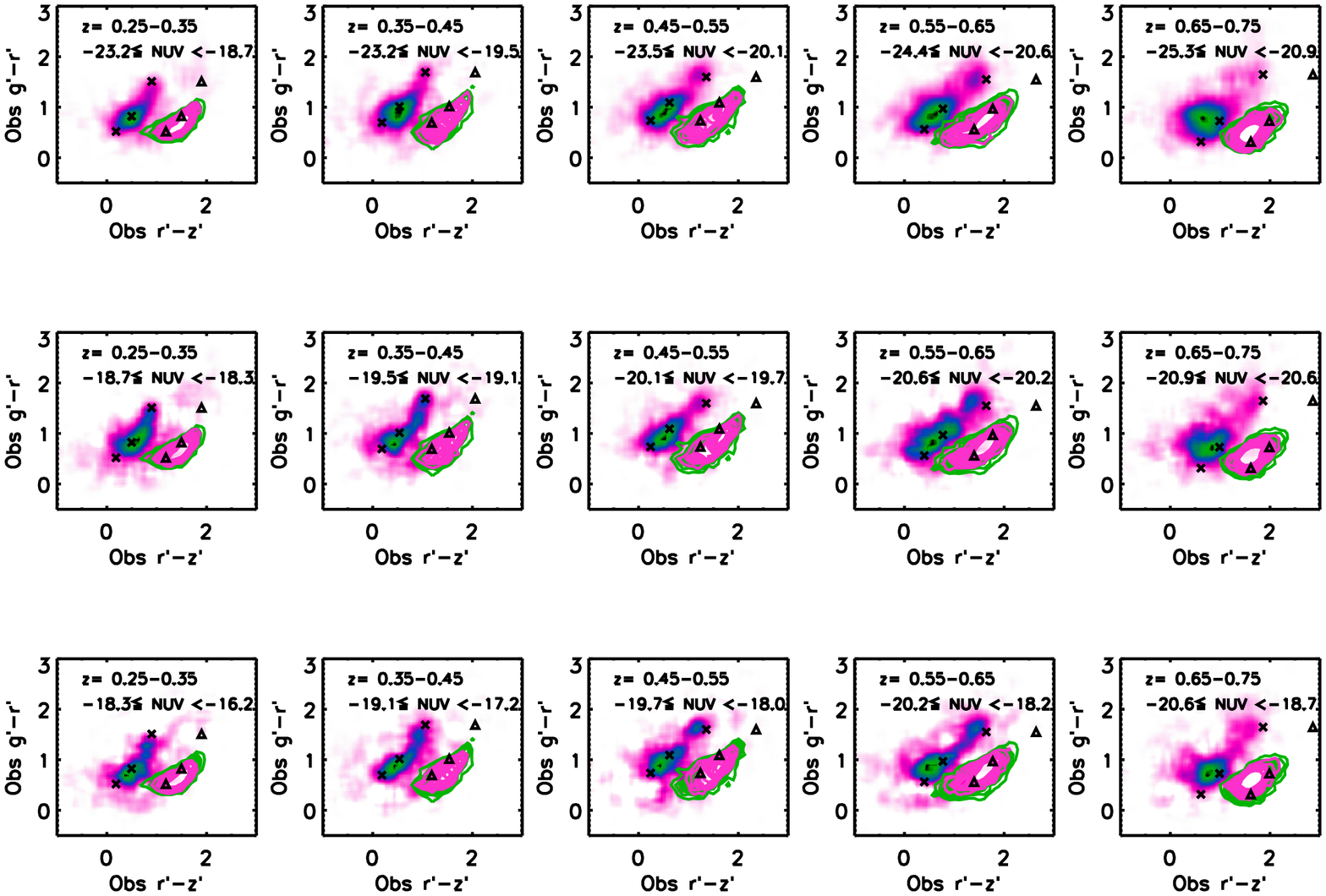} 
\caption{Similar to Fig. \ref{fig:ccplot_O2} but controlling for the markers' NUV luminosity.
At a fixed redshift bin, both the colors of the neighbors and the markers themselves remain similar regardless of the NUV luminosity.
\label{fig:ccplot_nuv}}
\end{figure*}

Since one of the primary WiggleZ target selection criteria
is based on UV flux, we also check whether the $NUV$ luminosity of 
the markers affects the color properties of the neighbors.  
This is essentially testing whether the total star-formation rate of 
the markers affects the color-color distributions of the neighboring galaxies.
To do so, we divide the markers into three groups in each redshift bin based on their rest-frame $NUV$ luminosity.
Because the sample of the markers is not complete to the same rest-frame $NUV$ depth, direct comparisons between different redshift bins cannot be done.
However, within the same redshift bin, we can compare the color-color distributions of the markers and their neighbors over its range of rest-frame $NUV$ luminosity.
The result is presented in Fig. \ref{fig:ccplot_nuv}.
As in the case for EW([OII]$\lambda$3727\AA),
we observe, and confirm with Kolmogorov-Smirnov tests, that within a fixed redshift bin the color distributions of the neighbors around markers with different $NUV$ luminosities are statistically identical. 
This supports our conclusion of Fig. \ref{fig:ccplot_O2} that the properties of the neighbors are not significantly affected by, or strongly correlated with,
 the characteristics of the markers.
We also find that the optical color distributions of the markers themselves do not strongly depend on the absolute $NUV$ luminosity.

\subsubsection{Control for AGN Candidates in the Markers}\label{sec:obsccm_agn}
\begin{figure}
\includegraphics[scale=0.36,angle=90]{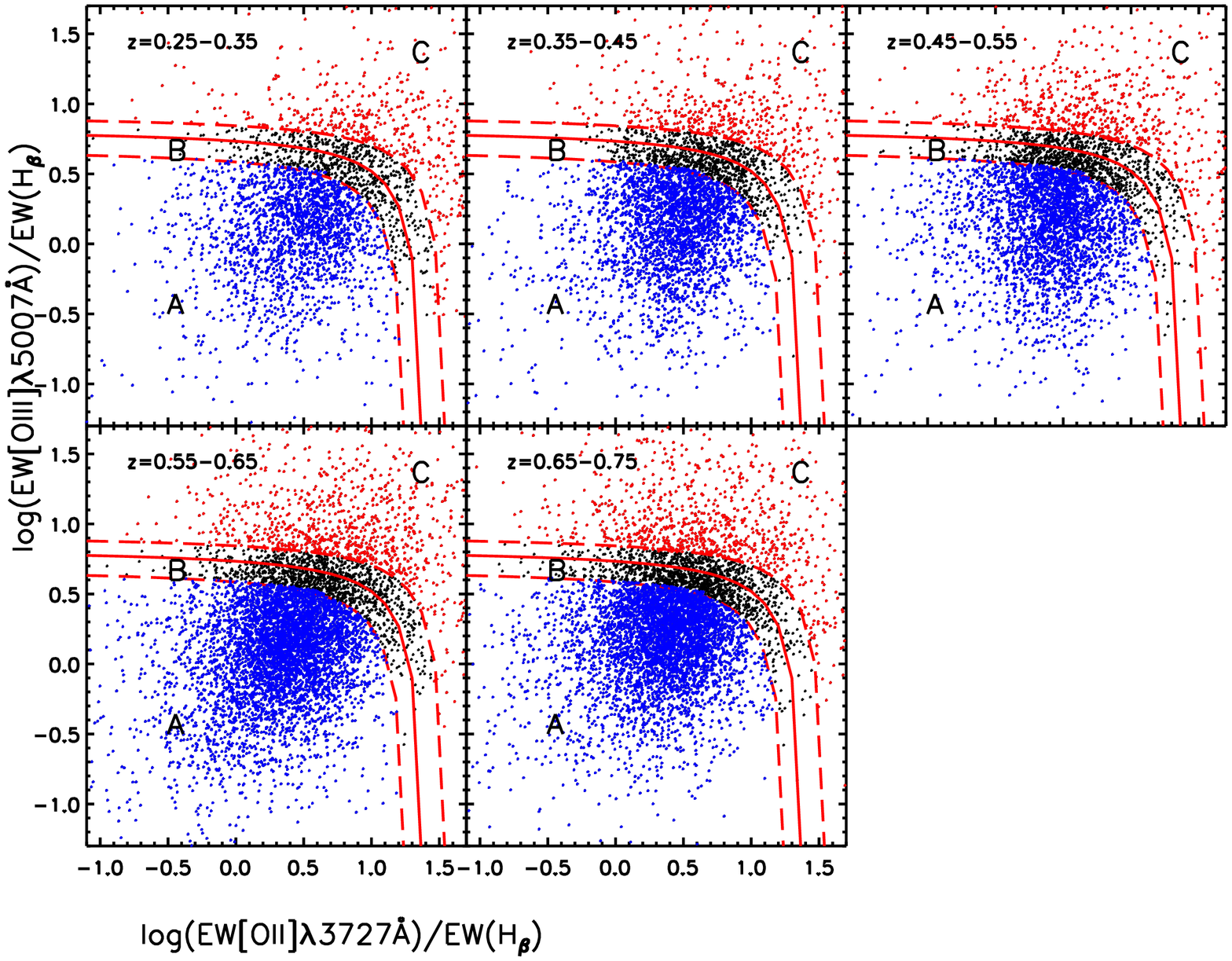}
\caption{Selection of AGN candidates. The WiggleZ galaxies are classified into three groups by the empirical separation used in \citet{2010A&A...510A..56B}.
The galaxies in group A are primarily star-forming galaxies, while those in group C are possibly AGN hosts. Group B contains all the rest of the galaxies between group A and C, within the regions enclosed by two dashed red lines in the plot
.
\label{fig:AGN}}
\end{figure}

Since AGN activities in galaxies are sometime responsible for emission lines, we are also interested in whether the neighbors of the AGN hosts have similar properties as those of normal star-forming galaxies, i.e., the rest of the WiggleZ galaxies.
The most common way to distinguish AGNs and star-forming galaxies is based on the ratios of [OIII]/$H_{\beta}$ and [NII]/$H_{\alpha}$ emission lines, the so-called BPT plot \citep{1981PASP...93....5B}.
This method works only at $z<0.48$ for the WiggleZ survey where all these emission lines 
can be detected within the spectral wavelength coverage.
Recently \citet{2010A&A...510A..56B} use a method similar to the BPT plot to separate AGNs and star-forming galaxies at $0.50 \leq z \leq 0.92$,
i.e., [OIII]$\lambda5007$\AA$/H_{\beta}$ versus [OII]/$H_{\beta}$,
 in the zCOSMOS survey.
The separation in this diagnostic diagram was derived empirically using the observed data by studying the positions in the diagram of AGN and star-forming galaxies which were classified based on the BPT plot \citep[e.g.,][]{1997MNRAS.289..419R,2004MNRAS.350..396L}.
We present the [OII]/$H_{\beta}$ vs [OIII]/$H_\beta$
 plot for all our WiggleZ galaxies at $0.25 \leq z \leq 0.75$ in Fig. \ref{fig:AGN}, 
where we overplot the analytical expression of Eq.~3 of \citet{2010A&A...510A..56B} for the demarcation curves between star-forming galaxies and AGNs.
We divide the WiggleZ markers into three groups based on their locations on the [OII]/$H_{\beta}$ plot.
Group A is those located below the analytical expression in the star-forming region. 
Group B contains a mix of star-forming and AGN galaxies, located in a narrow strip region centered at the analytical expression. 
Group C is the AGN candidates lying above the analytical expression.
These groups contribute about 67\%, 20\%, and 13\% of the galaxies, respectively.

\begin{figure}
\includegraphics[scale=0.37, angle=90]{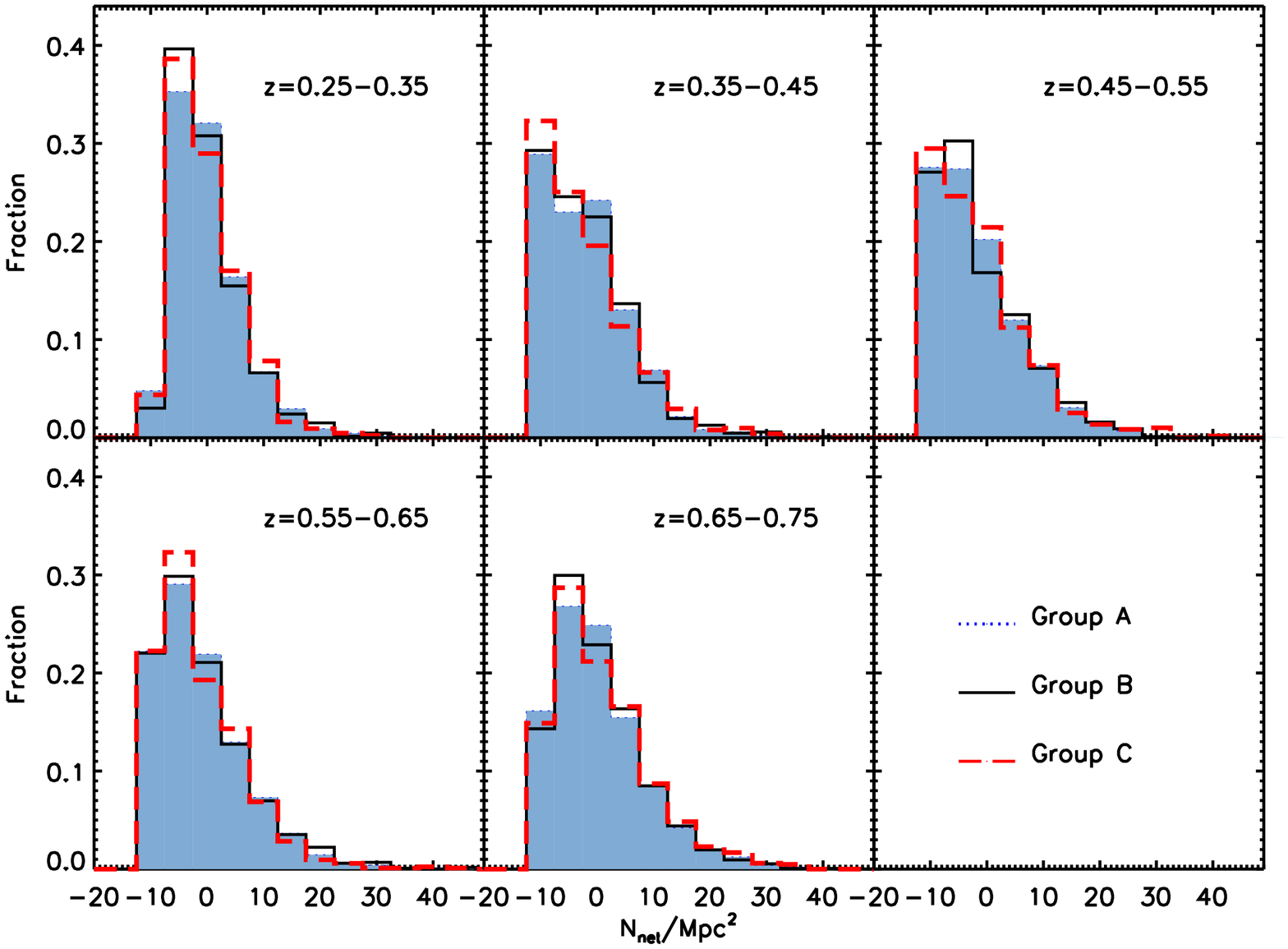}
\caption{
The $N_{net}/Mpc^2$ distributions of each group in Fig. \ref{fig:AGN}, presented as filled blue, thin black, and thick red histograms, respectively. The $N_{net}$ is computed using the $M_{r'}^*+1$ limit. Each histogram has a binsize of 5. All these these groups have similar $N_{net}/Mpc^2$ distributions at each redshift.
\label{fig:r1}}
\end{figure}

We find that the neighbors of each group have similar $N_{net}$ distribution (Fig. \ref{fig:r1}), 
and occupy essentially identical regions on the color-color diagram.
The $N_{net}$ here is computed using a limit of $M_{r'}^*+1$ , with $M_{r'}^*$ derived in \S\ref{subsec:LF}.
This echoes our conclusion that the properties of the neighbors are not strongly affected by the properties of the markers themselves.
Our results suggest that WiggleZ galaxies hosting AGN
are in environments similar to other WiggleZ galaxies.
This conclusion is consistent with other more detailed studies of 
Seyfert galaxies that the average environment of their hosts is not 
significantly different from other galaxies of similar properties 
\citep[e.g.,][]{1998ApJ...496...93D, 2001AJ....122.2243S}.

\subsection{Rest-Frame Color-Magnitude Diagrams \label{subsec:CMR}}
\begin{figure*}
\includegraphics[scale=0.74,angle=90]{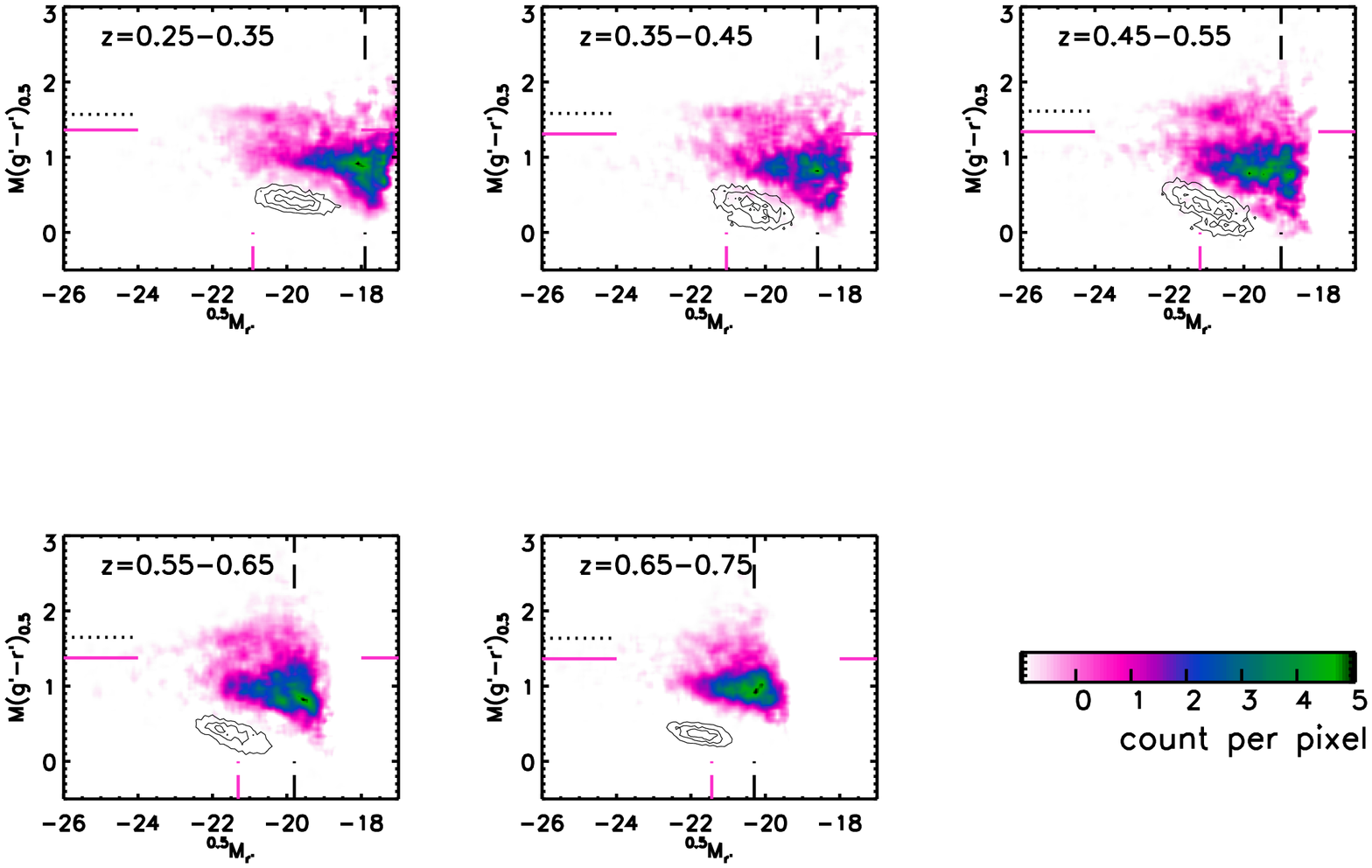}
\caption{The rest-frame color-magnitude of the WiggleZ neighbors at each redshift. The intensity of each panel is normalized to a total count of 1$\times10^4$,
and is in units of counts per pixel with a size of 0.0125$\times$0.025
(color$\times$magnitude) mag.
 The vertical pink dashed lines mark the derived $^{0.5}M_{r'}^*$ for the `All' subsample in \S\ref{subsec:LF}.
The horizontal solid line in each panel marks the separation between blue and red galaxies.
The color of the red sequence is indicated by the horizontal dotted line.
The black vertical dashed lines mark the $^{0.5}M_{r'}$ limit.
The red sequence of the neighbor galaxies is observable at each redshift.
For reference, the WiggleZ galaxies themselves are overplotted as the green contours shifted by --0.5 magnitudes in $M(g'-r')_{0.5}$.
\label{fig:abscmr}}
\end{figure*}

Figure \ref{fig:abscmr} presents the rest-frame color-magnitude diagrams for 
the WiggleZ neighbors in redshift bins of $z\sim$0.3 to $z\sim$0.7.
These are made by summing the elements of the color-color-magnitude cubes along the $x$-axis (i.e., $r'$-$z'$ axis).
Based on the conclusion in the previous two subsections (\S\ref{sec:obsccm_line} and \S\ref{sec:obsccm_agn}) that the properties of the neighbors are not strongly dependent on the properties of the WiggleZ galaxies themselves, 
the comparisons of the color-magnitude diagrams at difference redshifts can 
provide us useful insights into galaxy evolution.

The first observation we can glean from the Figure is that the WiggleZ neighbors populate 
regions of both the red sequence and the blue cloud at each redshift bin.
The majority of them are in the blue cloud.
A gap between the red sequence and blue cloud is seen.
In general, the red sequence at each redshift can be approximated by a
 horizontal line of the color of early-type galaxies.
The flatness of the red-sequence is more likely a result of
the relatively low signal of the red-sequence galaxies in our data, making
them insufficient for deriving an accurate fit, rather than due to the 
nature of the red sequence.
The dispersion in the red sequence appears to be larger at higher redshift; 
however, this can be mostly attributed to the larger photometric 
uncertainties for galaxies in these subsamples.
We also plot on Figure \ref{fig:abscmr} the color-magnitude distribution
of the WiggleZ galaxies in each redshift bin as contours, shifted by --0.5 mag 
in $g'-r'$ color. 
We note that the WiggleZ galaxies are primarily distributed along the
bright blue edge of blue-cloud galaxies, reflecting that they are 
strong star-forming galaxies.


\subsection{Galaxy Luminosity Function} \label{subsec:LF}
The  galaxy luminosity function (GLF) offers a convenient tool in exploring
the different components of the galaxy population in a sample.
The most widely used form for the GLF is the Schechter function \citep{1976ApJ...203..297S},
which can be characterized by three parameters: the normalization density $\phi^*$, the characteristic magnitude $M^*$, and the faint-end slope $\alpha$.
It has been found that $\alpha$ depends strongly on the galaxy SED type. 
The redshift evolution of the GLF, however, is also strongly dependent on SED types \citep[e.g.,][]{2003A&A...401...73W,2008ApJ...672..198L,2008A&A...477..763S}.
The GLF for early-type galaxies are described better with a shallower (sometime, down-turning) $\alpha$, and they are more abundant towards low redshift.
In contrast, late-type galaxies have a GLF with a steep $\alpha$, and their number density is largely unchanged toward low redshift.
In this subsection, we explore the GLF  for the WiggleZ neighbors,
and investigate the galaxy population components by examining 
the shape of their GLF and their evolution.

\subsubsection{Constructing the GLF \label{subsubsec:GLF}} 
The GLF of the neighbors is constructed by projecting the CCM cube along the $z$-axis (i.e., the $r'$ magnitude) to produce counts as a function of $r'$
magnitude.
We have conducted and cross-checked the analyses using both the observed and rest-frame CCM cubes. 
Here, we present only the results using the rest-frame CCM cubes, for which $g'$, $r'$, and $z'$ have been k-corrected to $z$=0.50.
We note that the observed $r'$-band at $z=0.5$ is approximately equivalent to
the rest $B$ band.

We also separate the neighbors into red and blue populations.
The division between the red and blue populations is chosen to be the 
$g'-r'$ color half way between the non-evolving early-type and the 
star-forming $\tau$ model of Figure \ref{fig:color_redshift} at each redshift, 
equivalent to 0.27 mag bluer than the red-sequence color.
To adjust for minor systematic effects in the photometry, the $g'-r'$
color (at rest $z=0.5$) of the red-sequence in each redshift bin in Fig. \ref{fig:abscmr}~is determined empirically as the reference.
This is done by examining the $g'-r'$ color distribution of galaxies in
each redshift bin which are brighter than $^{0.5}M_{r'}^*+1$ 
and have colors redder
than $(g'-r')_{rs}-0.27$, where $(g'-r')_{rs}$ is the non-evolving rest-frame model early-type galaxy color of Figure \ref{fig:color_redshift}.
The distribution is fitted with a Gaussian, and the peak is used as
the red-sequence color.
The red-sequence colors and the boundaries between the red and blue
populations are indicated in Fig. \ref{fig:abscmr} by the
dotted and solid  horizontal lines, respectively.
We note that the computed $g'-r'$ red-sequence colors are essentially identical
to those of the models, with the exception of the $z=0.3$ bin, 
where the $g'-r'$ separation for red and blue galaxies appears to be $\sim$0.06 bluer than the computed color of $(g'-r')_{rs}-0.27$. 

\begin{figure*}
\includegraphics[scale=0.74,angle=90]{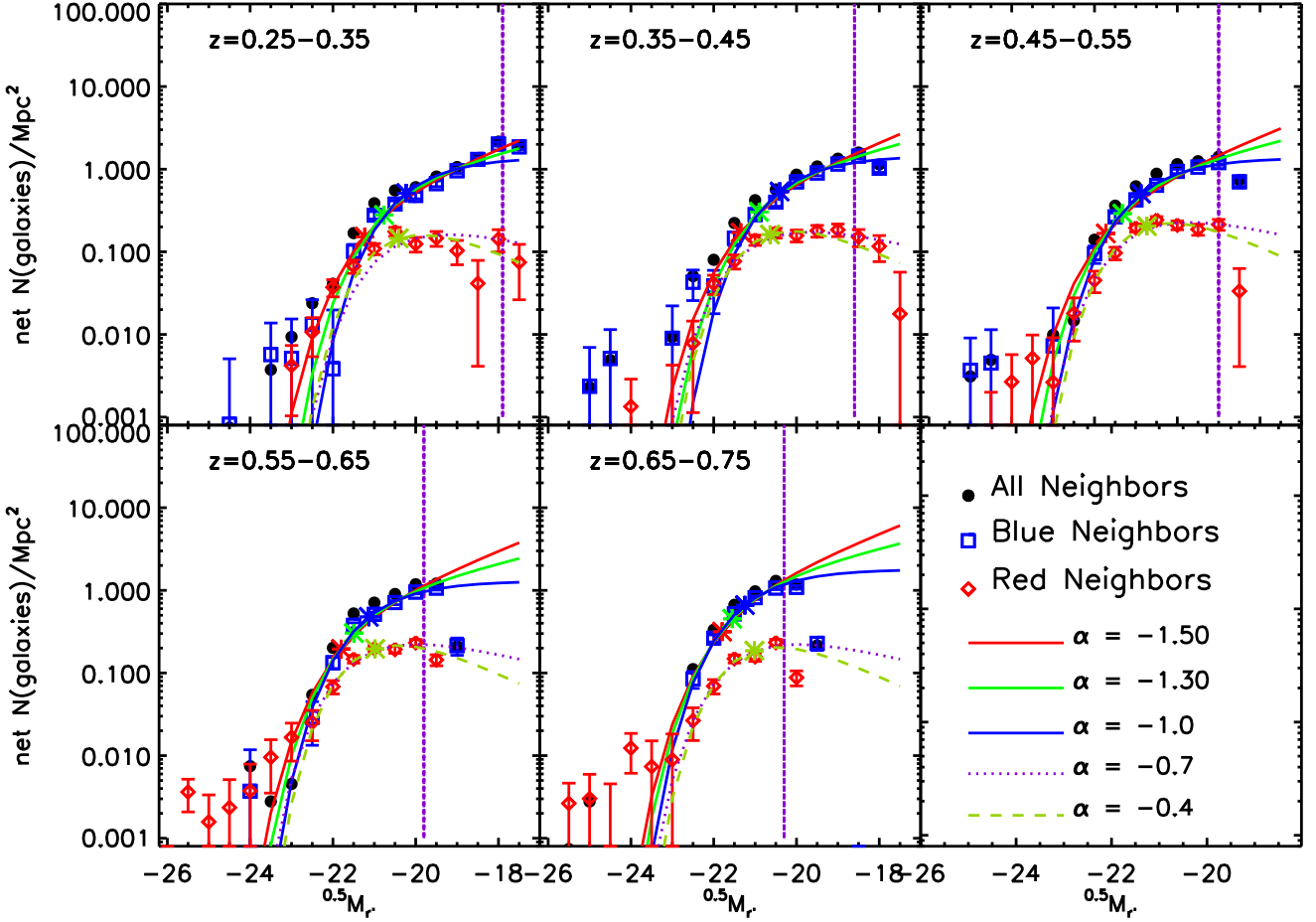}
\caption{The $r'$-band luminosity function for all the neighbors of the WiggleZ galaxies.
The red, green, and blue curves are the fitted Schechter LF with $\alpha$=-1.50, 1.30, and -1.0, respectively.
The Schechter LF for the red neighbors are described by $\alpha$= -0.70 and -0.40 shown as the dotted purple and dashed green curves.
The vertical purple line marks the sample complete limit. 
The GLF for the `All' and `Blue' neighbors are better described by a steeper $\alpha$ than that to the `Red' subsample.
\label{fig:LF}}
\end{figure*}

\begin{figure}
\includegraphics[scale=0.37,angle=90]{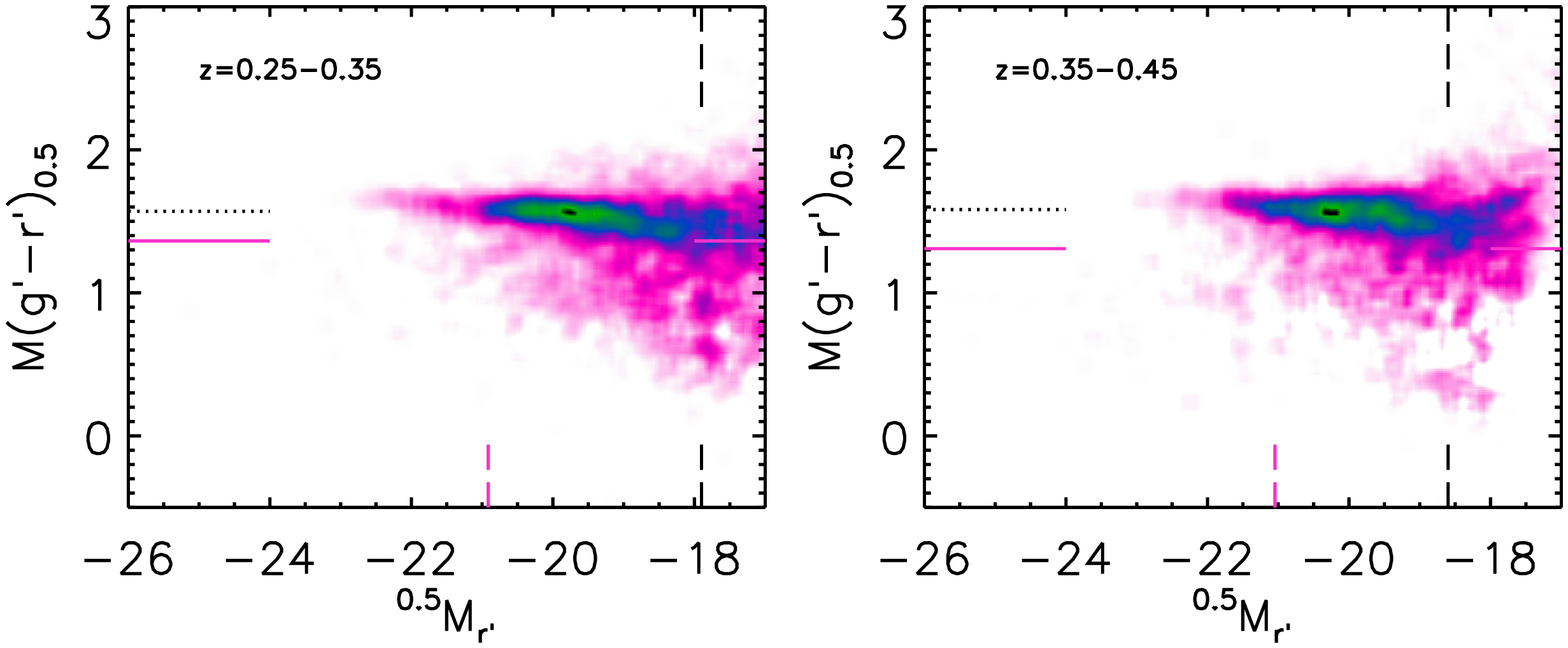}
\includegraphics[scale=0.37,angle=90]{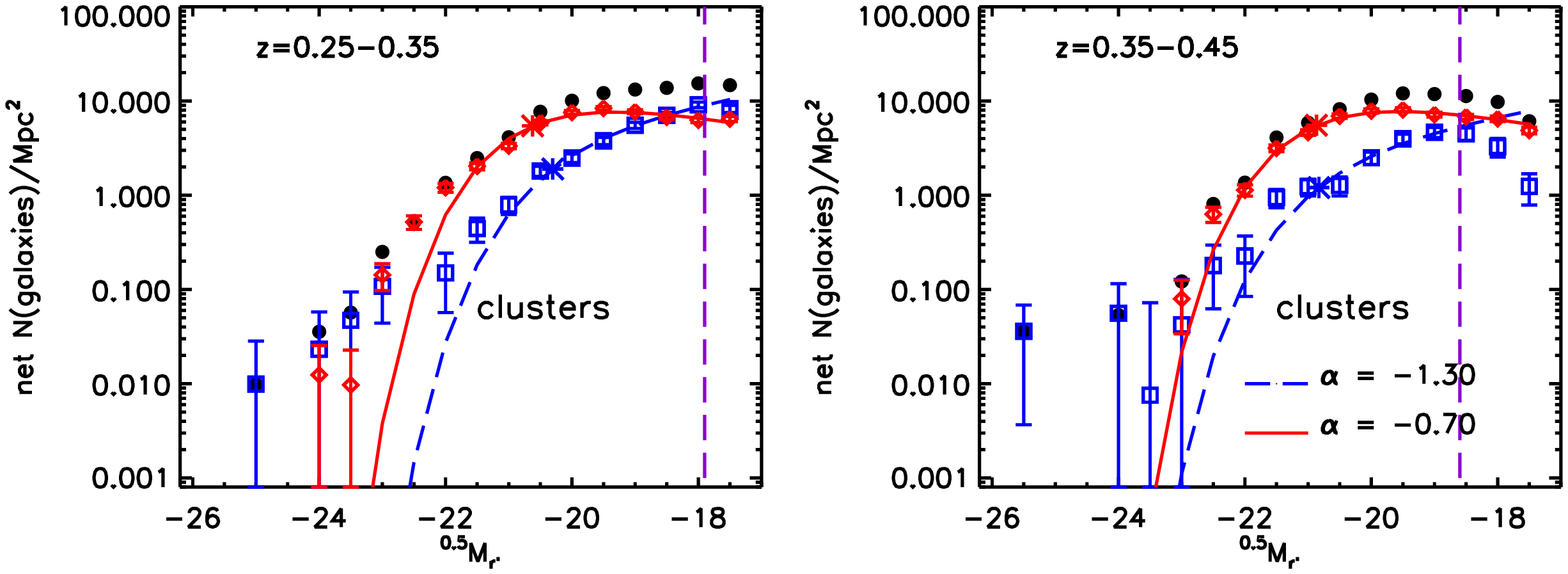}\\
\caption{The color-magnitude diagrams (Top) and luminosity functions (Bottom) for the RCS-WSF sample for the lowest two redshift bins. The red-sequence galaxies are dominant in both the plots. The symbols are of the same meaning as Fig. \ref{fig:LF}.
\label{fig:LF_cl}}
\end{figure}

The GLF results are presented in Figure \ref{fig:LF}. 
We denote the three subsamples of galaxies and their GLF
 as `All', `Blue', and `Red'.
The errors in the y-axis in each $r'$ bin is computed using  Poisson statistics.
We also compute the GLF using the same method for the RCS-WSF sample for 
comparison with the WiggleZ neighbor galaxies at the two lower redshift bins,
allowing us to examine the effects of environment on galaxy evolution.
The CMD and the resultant GLFs for the cluster sample are
plotted in Fig. \ref{fig:LF_cl}. 


\subsubsection{The Schechter Function Fit}

We fit the GLF discussed above using the Schechter function.
We determine the completeness limit in $^{0.5}M_{r'}$~for each redshift bin by examining
the total net counts in 0.1 magnitude bins as a function of $^{0.5}M_{r'}$, smoothed
by a three-bin kernel. 
We use the bin 0.1 mag brighter than the peak as the limit for fitting
the Schechter function, giving
$^{0.5}M_{lim}=$ [--17.9, --18.6, --19.0, --19.8, --20.3] for the
five redshift bins. 
These limits are marked as vertical purple (dashed) lines in Fig. \ref{fig:LF}.

To investigate the shape of the GLF, 
we first allow $\alpha$ to vary in fitting the  Schechter function.
The results of the Schechter function fits are tabulated in Table \ref{tab:LF}.
As evident from the CMD (Fig.~\ref{fig:abscmr}), 
the GLFs in Fig. \ref{fig:LF} of the WiggleZ neighbor galaxy populations are dominated by blue galaxies.
It is not surprising that the GLFs of the `All' and the `Blue' subsamples
are similar, because $\sim80$\% of the neighbors are blue galaxies.

For the `Blue' and 'All' samples, we find that the fitted $\alpha$ appear to 
become less steep at higher redshift.
However, this could in part be the result of the different sampling magnitudes
in the subsamples; the shallower absolute magnitude limits in the 
higher redshift bins make $\alpha$ less well-determined.
This is especially true at the highest redshift bin, where the completeness
limit is only about 1 to 1.5 mag past $M^*$.
Discarding the $z=0.7$ bin, 
the  derived $\alpha$ for the `All' and `Blue' subsamples range between $\sim-1.0$ and $\sim-1.5$, comparable to those in the literature for star-forming or late-type galaxies \citep[e.g.,][]{2003A&A...401...73W,2009MNRAS.400..429C, 2008ApJ...672..198L}.
  The possibility that the apparent changes in $\alpha$ at different 
redshifts are due to the change in the sampling limit can be demonstrated 
by fitting the three low-redshift bins for the `Blue' subsample
to a common limit 
of $^{0.5}M_{r'}= -19.0$, which is complete for the three bins.
We obtain the results: 
$\alpha=$ [$-1.15\pm0.17$, $-1.38\pm0.16$, $-1.03\pm0.12$]
for $z=$ [0.3, 0.4, 0.5], which are consistent 
with being identical at $\sim 1.75\sigma$.

We overplot the Schechter LF in Fig. \ref{fig:LF} with fixed $\alpha$=[-1.50, -1.30, -1.00] fitted for the `Blue' subsamples to show the ability of the
data to distinguish between the faint-end slopes within this range.
By combining the three lower-redshift bins ($0.25<z<0.55$) where the data
are of sufficient depth to obtain a robust measurement of $\alpha$, 
 we find for the `Blue' galaxies a best-fit $\alpha$ of $-1.18\pm0.08$. 
For consistency of comparison of $^{0.5}M_{r'}$ among redshift bins, we 
adopt an $\alpha$ of $-1.3$ for all redshift bins, and tabulate the 
fitted $^{0.5}M_{r'}^*$ in Table \ref{tab:LF}.

The best fitting results for $^{0.5}M_{r'}^*$ and $\alpha$ for the WiggleZ
`Red' neighbor galaxy samples at different redshift bins are also presented 
in Table \ref{tab:LF}.
For the lower redshift bins, where the data are of sufficient depth, 
the GLF is considerably less steep 
at the faint-end compared to that of the `Blue' subsample. 
At higher redshifts, the red GLFs can be described by a steeper $\alpha$, 
but this is again likely because the faint-end of the GLF is not observable at these redshifts.
We use the three lower redshift bins to establish 
a more robust estimate of the faint-end slope of the red galaxy GLF.
We obtain $\alpha = -0.45\pm0.13$ by summing the data (to $^{0.5}M_{r'}=-19.0$) in these bins. 

For the purpose of comparison, we perform the same analysis 
 for the RCS-WSF sample of markers in RCS2 clusters. 
Here, the CMD shows a very strong red sequence and the
total GLF is dominated by red galaxies, as shown in Figure \ref{fig:LF_cl}).  
The faint-ends of the `Blue' galaxy GLF in the two redshift bins
have a similar $\alpha$ as those for the corresponding subsamples of the WiggleZ neighbors.
For the purpose of comparing $M^*$, we also fit the RCS-WSF `Blue' samples
using $\alpha$ = [-1.50, -1.30, -1.00] with the results listed in
Table \ref{tab:LF_cl}.

Opposite to the WiggleZ neighbor samples, the GLF of the RCS-WSF sample
of cluster galaxy targets is dominated by the `Red' galaxy sample.
The faint-end slope for these red galaxies appears to be significantly steeper 
than that of the WiggleZ counterpart, with $\alpha$ steeper than --0.65.
The combined `Red' GLF for the two redshift bins of the RCS-WSF data
produces a best fitting $\alpha=-0.74\pm0.06$. 
A similar combination for the WiggleZ neighbor sample produces 
$\alpha=-0.18\pm0.17$. 
In order to provide a consistent comparison for the $^{0.5}M_{r'}^*$
values for the `Red' WiggleZ neighbor sample and the `Red' RCS2 cluster 
neighbor sample, we also refit all the red galaxy subsamples using $\alpha=-0.4$
and --0.7, and the results are tabulated in Table \ref{tab:LF}.

\subsubsection{The Evolution of $^{0.5}M_{r'}^*$} \label{subsubsec:Q}

\begin{figure}
\includegraphics[scale=0.37,angle=90]{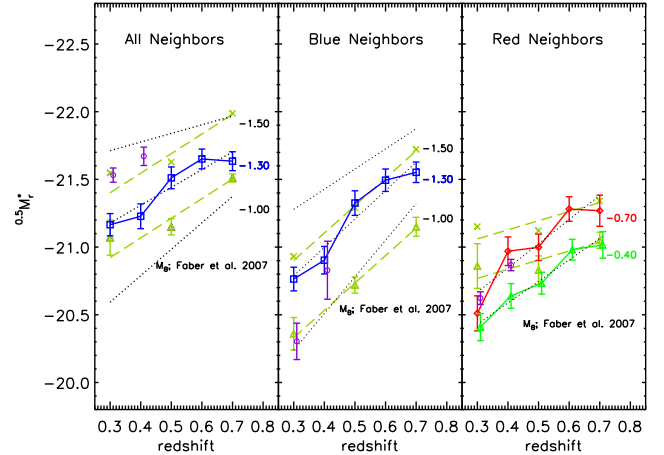}
\caption{$^{0.5}M_{r'}^*$ as a function of redshift derived from Fig. \ref{fig:LF} using three fixed $\alpha$.
The dashed curve with triangles is the $M_B^*$ from Faber et al. (2007), and the one with crosses symbols is the $^{0.5}M_{r'}$ by applying a simple color transfer from $B$ to $r'$. The linear fits for the WiggleZ neighbor galaxies are overplotted as the dotted lines. 
The RCS-WSF sample is overplotted as purple open circle.
The derived evolutionary term, 
i.e., the slope of the fit, is --1.31 for all the neighbors (the left panel), --2.10 with $\alpha$= --1.30 for the blue neighbors (the middle panel), and --1.59 for the red neighbors with $\alpha$=--0.40.
\label{fig:Q}}
\end{figure}

We examine the evolution of the $M^*_{r'}$ parameter of the Schechter LF 
by assuming a simple linear dependence between $M^*_{r'}$ and redshift,
as used by \citet{1999ApJ...518..533L} and others.
We can write:
$^{0.5}M_{r'}^*(z)$=$m_{r'}^*-DM-(k-k_{z=0.50})+Q(z-0.5)$, where $DM$ is the 
distance modulus and $k$ is the k-correction. 
The evolution term $Q$ is derived from a linear fit to $^{0.5}M_{r'}^*$ as a 
function of redshift in the form of $^{0.5}M_{r'}^*(z)$ = $^{0.5}M_{r'}^*(z=0.5) + Q(z-0.5)$.
The derived $Q$ depends on the value of $\alpha$;
a steeper $\alpha$ results in a smaller $Q$.
Because the fitted $\alpha$ and $M_{r'}^*$ correlate with each other and $\alpha$ are marginally different in our redshift divisions, 
we use the $M_{r'}^*$ derived with $\alpha= -1.30$ for the `All' and `Blue' subsamples and $\alpha$= --0.40 for the `Red' neighbors, 
obtaining $Q$=[$-1.31\pm0.24$,$-2.10\pm0.26$,$-1.59\pm0.30$] for the `All', `Blue', and `Red' subsamples, respectively.
The results are shown in Fig. \ref{fig:Q} and tabulated in Table \ref{tab:LF}.
We also derive $Q$ using $\alpha=-1.0$ and --1.5 for the `All' and `Blue' neighbors, and $\alpha=-0.7$ and --0.4 for the `Red' subsample.
These fits are shown in Fig. \ref{fig:Q} as dotted line for each redshift bins.
	\begin{table}
        \caption{Parameters for Luminosity Functions of the WiggleZ Neighbors \label{tab:LF}}
        \begin{tabular}{llll}
        \hline
        \bf All Neighbors \rm \\
        \hline\hline
         redshift  &  $^{0.5}M_{r'}^*$ & $\alpha$ \\
        \hline
        fitting $\alpha$ \\
0.25--0.35 & -21.46$\pm$0.18 & -1.42$\pm$0.05 \\
0.35--0.45 & -21.33$\pm$0.21 & -1.35$\pm$0.08 \\
0.45--0.55 & -20.89$\pm$0.14 & -0.89$\pm$0.10 \\
0.55--0.65 & -21.23$\pm$0.15 & -0.97$\pm$0.12 \\
0.65--0.75 & -21.10$\pm$0.16 & -0.76$\pm$0.18 \\
        \hline\hline
         redshift  &  $^{0.5}M_{r'}^*$ & $^{0.5}M_{r'}^*$ & $^{0.5}M_{r'}^*$ \\
          & $\alpha$= -1.50 & $\alpha$= -1.30 & $\alpha$= -1.00 \\
        \hline
0.25--0.35 & -21.68$\pm$0.11 & -21.16$\pm$0.08 & -20.60$\pm$0.06\\
0.35--0.45 & -21.70$\pm$0.12 & -21.22$\pm$0.09 & -20.69$\pm$0.06\\
0.45--0.55 & -21.92$\pm$0.10 & -21.51$\pm$0.08 & -21.02$\pm$0.05\\
0.55--0.65 & -21.98$\pm$0.09 & -21.65$\pm$0.07 & -21.26$\pm$0.05\\
0.65--0.75 & -21.89$\pm$0.08 & -21.63$\pm$0.06 & -21.30$\pm$0.05\\
        \hline
	Q   & -0.64$\pm$0.32  & -1.31$\pm$0.24 & -1.95$\pm$0.19 \\ 
        $^{0.5}M_{r'}^*(z=0.5)$ & -21.84$\pm$0.05 & -21.44$\pm$0.04 & -20.98$\pm$0.03 \\
        \hline\hline
        \hline
        \bf Blue Neighbors \\
        \hline\hline
         redshift  &  $^{0.5}M_{r'}^*$ & $\alpha$ \\
        \hline
        fitting $\alpha$ \\
0.25--0.35 & -21.41$\pm$0.25 & -1.56$\pm$0.06 \\
0.35--0.45 & -21.34$\pm$0.29 & -1.49$\pm$0.10 \\
0.45--0.55 & -20.90$\pm$0.17 & -1.02$\pm$0.11 \\
0.55--0.65 & -21.19$\pm$0.19 & -1.05$\pm$0.15 \\
0.65--0.75 & -21.02$\pm$0.18 & -0.74$\pm$0.21 \\
        \hline\hline
         redshift  &  $^{0.5}M_{r'}^*$ & $^{0.5}M_{r'}^*$ & $^{0.5}M_{r'}^*$ \\
          & $\alpha$= -1.50 & $\alpha$= -1.30 & $\alpha$= -1.00 \\
        \hline
0.25--0.35 & -21.23$\pm$0.11 & -20.76$\pm$0.08 & -20.23$\pm$0.06\\
0.35--0.45 & -21.34$\pm$0.13 & -20.90$\pm$0.10 & -20.40$\pm$0.07\\
0.45--0.55 & -21.72$\pm$0.11 & -21.32$\pm$0.09 & -20.87$\pm$0.06\\
0.55--0.65 & -21.80$\pm$0.10 & -21.49$\pm$0.08 & -21.13$\pm$0.06\\
0.65--0.75 & -21.79$\pm$0.09 & -21.55$\pm$0.07 & -21.24$\pm$0.06\\
        \hline
	Q  & -1.49$\pm$0.34 & -2.10$\pm$0.26 & -2.71$\pm$0.20 \\
        $^{0.5}M_{r'}^*(z=0.5)$ & -21.58$\pm$0.05 & -21.21$\pm$0.04 & -20.78$\pm
$0.03 \\
        \hline
        \bf Red Neighbors \\
        \hline\hline
         redshift  &  $^{0.5}M_{r'}^*$ & $\alpha$ \\
        \hline
        fitting $\alpha$ \\
        \hline
0.25--0.35 & -20.23$\pm$0.20 & 0.099$\pm$0.27 \\
0.35--0.45 & -20.74$\pm$0.22 & -0.50$\pm$0.19 \\
0.45--0.55 & -20.60$\pm$0.20 & -0.26$\pm$0.21 \\
0.55--0.65 & -21.21$\pm$0.24 & -0.64$\pm$0.22 \\
0.65--0.75 & -21.33$\pm$0.38 & -0.77$\pm$0.37 \\
        \hline\hline
         redshift  &  $^{0.5}M_{r'}^*$ \\
          & $\alpha$= -0.70 & $\alpha$= -0.40 \\
        \hline
0.25--0.35 & -20.51$\pm$0.12 & -20.40$\pm$0.09\\
0.35--0.45 & -20.96$\pm$0.10 & -20.63$\pm$0.09\\
0.45--0.55 & -20.99$\pm$0.09 & -20.73$\pm$0.08\\
0.55--0.65 & -21.28$\pm$0.09 & -20.98$\pm$0.07\\
0.65--0.75 & -21.26$\pm$0.11 & -21.01$\pm$0.09\\
        \hline
	Q  & -1.79$\pm$0.37  & -1.59$\pm$0.30\\
        $^{0.5}M_{r'}^*(z=0.5)$ & -21.02$\pm$0.05 & -20.76$\pm$0.04 \\
        \hline
        \end{tabular}
        \end{table}

        \begin{table}
        \caption{Parameters for Luminosity Functions of the RCS-WSF Sample \label{tab:LF_cl}}
        \begin{tabular}{llll}
        \hline
        \bf All Neighbors \rm \\
        \hline\hline
         redshift  &  $^{0.5}M_{r'}^*$ & $\alpha$ \\
        \hline
        fitting $\alpha$ \\
0.25--0.35 & -21.04$\pm$0.08 & -1.10$\pm$0.03 \\
0.35--0.45 & -21.05$\pm$0.10 & -0.94$\pm$0.06 \\
        \hline\hline
         redshift  &  $^{0.5}M_{r'}^*$ & $^{0.5}M_{r'}^*$ & $^{0.5}M_{r'}^*$ \\
          & $\alpha$= -1.50 & $\alpha$= -1.30 & $\alpha$= -1.00 \\
        \hline
0.25--0.35 & -22.14$\pm$0.07 & -21.53$\pm$0.05 & -20.81$\pm$0.03\\
0.35--0.45 & -22.14$\pm$0.09 & -21.67$\pm$0.06 & -21.13$\pm$0.05\\
        \hline\hline\hline
        \bf Blue Neighbors \rm \\
        \hline\hline
         redshift  &  $^{0.5}M_{r'}^*$ & $\alpha$ \\
        \hline
        fitting $\alpha$ \\
0.25--0.35 & -20.94$\pm$0.41 & -1.52$\pm$0.03 \\
0.35--0.45 & -21.59$\pm$0.73 & -1.55$\pm$0.06 \\
        \hline\hline
         redshift  &  $^{0.5}M_{r'}^*$ & $^{0.5}M_{r'}^*$ & $^{0.5}M_{r'}^*$ \\
          & $\alpha$= -1.50 & $\alpha$= -1.30 & $\alpha$= -1.00 \\
        \hline
0.25--0.35 & -20.87$\pm$0.18 & -20.30$\pm$0.13 & -20.00$\pm$-0.0\\
0.35--0.45 & -21.39$\pm$0.29 & -20.82$\pm$0.21 & -20.20$\pm$0.15\\
        \bf Red Neighbors \rm \\
        \hline\hline
         redshift  &  $^{0.5}M_{r'}^*$ & $\alpha$ \\
        \hline
        fitting $\alpha$ \\
0.25--0.35 & -20.52$\pm$0.09 & -0.64$\pm$0.06 \\
0.35--0.45 & -20.81$\pm$0.09 & -0.66$\pm$0.06 \\
        \hline\hline
         redshift  &  $^{0.5}M_{r'}^*$ \\
          & $\alpha$= -0.70 & $\alpha$= -0.40 \\
        \hline
0.25--0.35 & -20.62$\pm$0.04 & -20.14$\pm$0.04\\
0.35--0.45 & -20.86$\pm$0.04 & -20.45$\pm$0.03\\
        \hline
        \end{tabular}
        \end{table}


We adopt the parametrization of
$^{0.5}M_{r'}^*(z)= -21.44 - 1.31(z-0.5)$ from the `All' sample
in the derivation of the red-galaxy fraction in the following subsection.
 
\subsection{The Fraction of Red Neighbors} \label{subsec:fred}
\begin{figure*}
\includegraphics[scale=0.74,angle=90]{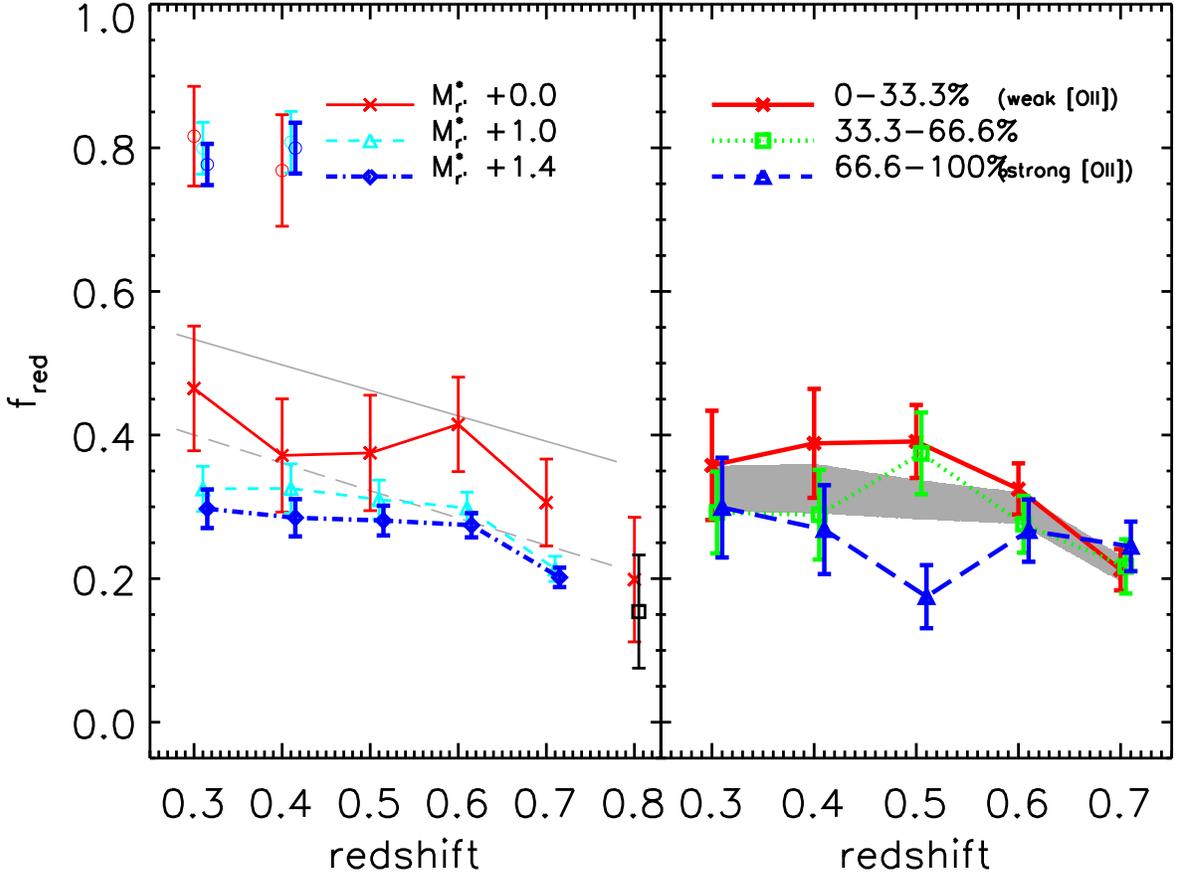}
\caption{Left: Fraction of red neighbors $f_{red}$ as a function of redshift to different $M_{r'}$ depth. Different curves represent the $f_{red}$ computed using different magnitude depth as noted in the plot. 
The gray straight lines are from \citet{2010A&A...509A..40I} for their Sample III ($< M^*+0.8$) `All' (solid linestyle) and `Isolated' (dashed linestyle) subsamples. 
The $f_{red}$ of the RCS-WSF sample at $z\sim$0.30 and 0.40 are overplotted as individual open symbols, which have a value of $\sim$0.8.
The $f_{red}$ for the $z=0.75-0.85$, which is beyond the redshift range 
of our main sample, are plotted as individual points for magnitude limits
of $M^*_{r'}$ and $M^*_{r'}+1.2$ (black square) to illustrate the 
continuous decrease of \fred~beyond $z\sim0.7$.
Right: $f_{red}$ computed using $^{0.5}M_{r'}^*+1$ depth after controlling for the EW([OII]$\lambda$3727\AA) of the WiggleZ markers. The gray shaded region is the $f_{red}$ using all markers with the same magnitude limit.
The $f_{red}$ is similar regardless of the properties of the markers.
\label{fig:fred}}
\end{figure*}

In Figure \ref{fig:obsccplot} we have observed that dusty star-forming 
galaxies are not common for both the WiggleZ galaxies and their neighbors.
The red neighbors of the WiggleZ galaxies are more likely to be those 
which have completed their star formation.
We investigate the fraction of these red passive neighbors ($f_{red}$) as a function of redshift.
We define red neighbor galaxies the same way as described in \S\ref{subsubsec:GLF}.
The boundaries separating the blue and the red galaxies are shown in Fig. \ref{fig:abscmr}, which is about 0.27 mag bluer than the red-sequence, k-corrected to $z=0.5$.
This color separation is equivalent to the gap in the galaxy bimodal color distribution at a fixed redshift seen in our samples.
The fraction of net red neighbors to the total net neighbor counts
 as a function of redshift is presented in Fig. \ref{fig:fred}. 
We compute \fred~to four different depths of 
$^{0.5}M_{r'}^*+\Delta M$ with $^{0.5}M_{r}^*$ as described
 in \S\ref{subsubsec:Q} and 
$\Delta M$=[0.0, 0.5, 1.0, 1.4], 
with the largest $\Delta M$ determined by
the depth of sampling of the data from the largest redshift bin.
The error in $f_{red}$ is estimated using Poisson statistics.

We observe that $f_{red}$ is not a strong function of redshift. 
Within the uncertainties of the measurements, \fred~can be described
as flat between $z=0.25$ and 0.65, with an average of 
 $\sim$0.28$\pm$0.01 
for the magnitude limit of $^{0.5}M_{r'}^*+1.4$ .
At $z\sim0.7$, there appears to be a drop in
\fred~at $\sim 3.5\sigma$ level.
We also see a relatively small change in \fred~within the relatively
 magnitude range we probed -- changing by the order 0.1 over the 
1.4 magnitude range, often within the uncertainties of the measurements
in the same redshift bin.
However, we note that differentially, as indicated by the GLFs of
the red and blue galaxies, the fraction of blue galaxies
increases significantly at the faint-end.
Using data combining the two lowest redshift bins 
where we can sample down to $^{0.5}M_{r'}^*+2.5$, we find 
\fred$\sim0.22\pm0.01$, a difference of $\sim5\sigma$
from the \fred$\sim0.45\pm0.04$ measured at $^{0.5}M_{r'}^*$.

For comparison, the $f_{red}$ for galaxies around the RCS-WSF sample of cluster galaxy markers are also plotted in Fig. \ref{fig:fred}, which have an average $f_{red}\sim0.8$ for the $z\sim$0.3 and 0.4 bins.
As expected, the $f_{red}$ for the RCS-WSF sample is much larger than that for the WiggleZ neighbors.

We also compute the $f_{red}$ values for subsamples of WiggleZ markers in
different EW([OII]$\lambda$3727\AA) bins.
We divide the markers into three groups based on the 33.3 and 66.7 percentiles
 in the distribution of EW([OII]$\lambda$3727\AA).
The $f_{red}$ values, computed using the limit of $^{0.5}M_{r'}^*+1$, as a 
function of redshift are presented in the right panel of Figure. \ref{fig:fred}.
The gray shaded area is the $f_{red}$ calculated with the same magnitude limit using all the neighbors (i.e., the dashed blue curve in the left panel), overlaid for comparison.
Although the WiggleZ markers with stronger EW([OII]$\lambda$3727\AA) appear
to have somewhat lower $f_{red}$, the $f_{red}$ for the different bins
are well within the individual measurement uncertainties, with the exception
of the $z=0.5$ bin.
Averaging over all redshift bins, we find the mean $f_{red}$ for 
the three EW([OII]$\lambda$3727\AA) bins, from low to high, to be 
$0.305\pm0.019$, $0.278\pm0.021$, and $0.246\pm0.021$, 
or, a difference of
about 2$\sigma$ between the weakest and strongest [OII]$\lambda$3727\AA~samples.
Thus, there is some evidence that the neighbors of markers with the largest
EW([OII]$\lambda$3727\AA~may have a lower average \fred, indicating regions
around galaxies with large specific star-formation rates may have a larger
fraction of star-formation galaxies.
However, the relatively low significance difference (which comes mostly
from the $z=0.5$ bin) is consistent with our 
conclusion of Figures \ref{fig:ccplot_O2} and \ref{fig:ccplot_nuv} that the properties of the neighbors are not strongly dependent on the 
characteristics of the WiggleZ galaxies themselves.

\section{Discussion} \label{sec:discussion}
\subsection{Robustness of Our Results}
We have demonstrated that our novel approach -- probing photometric properties
of galaxies using color-color-magnitude cubes of neighbors around galaxies
of known redshift -- has yielded robust and interesting results.
The advantages of our method are: (1) being straightforward, as it is 
equivalent to counting galaxies within an aperture and applying statistical
background corrections;  (2) not sensitive to whether the sample 
of the markers is complete; 
(3) allowing us to derive the photometric properties of galaxies to a limit
considerably deeper than the corresponding limits for the spectroscopic 
sample; and (4) providing a complete census of the neighboring galaxies 
independent of the spectral/color properties of the galaxies.
The key point in the method is the assumption that the marker galaxies and 
the excess galaxy counts
around them are at the same redshift.
The net neighbor counts around an individual galaxy are not statistically
 meaningful, 
but stacking color-color-magnitude cubes of a large number of markers provide 
statistically significant quantities.
However, much care is required in carrying out the procedure.

First, proper background correction is essential in our method, 
especially when the signal of the net excess is only a fraction of the total 
galaxy counts.
Any systematic discrepancy in the background subtraction would have a profound
effect on the result.
If the background is under-subtracted, the intrinsic properties of the net excess 
galaxies will be overwhelmed by the background counts; for instance, it may 
result in a power-law galaxy luminosity function without an apparent bend/knee.
On the other hand, over-subtracting the background may result in small, or
even negative, net counts, and hence no analyses can be done.
The very large angular size of the RCS2 patches allows us to use 
photometric data from the same patch as the markers to estimate the 
background correction, minimizing any possible systematic issues.
The use of uniform random catalogs to map out the imaging survey area,
CCD chip gaps, bad CCD columns, bright star halos and other artifacts, ensure
the proper treatment of the sampling aperture size.
Our exercise of measuring excess counts centered on a large number of random
positions in \S\ref{sec:rcs2} demonstrates that we have handled the background correction properly, as the net excess around the randomly drawn points has a mean value of essentially zero and is not a function of galaxy magnitudes or colors.

Another way of verifying the robustness of our background subtraction
technique is to use the WiggleZ redshift sample.
While it poorly samples the whole photometric galaxy catalog, the very large
sample of WiggleZ redshifts allows us to test the principle of
background subtraction in general, and our technique in particular.
We perform this test by comparing the ratio of the net excess galaxy counts 
to the total counts within the $r_p=0.25$Mpc aperture in both the 
photometric and WiggleZ redshift samples.
We note that this method may not produce an exact comparison, 
as the WiggleZ sample has a number of selection criteria which may 
produce different selection effects for galaxies at different redshifts 
that are not possible to mimic using a purely magnitude-limited 
photometric sample.  
The most significant selection effect that produces a significant
difference in the redshift distribution between a $r'$-band selected
sample and the WiggleZ sample is the low-redshift-rejection (LRR) 
applied to the selection of WiggleZ targets (see \S2.3).
Thus, we limit our comparison to using data from WiggleZ markers in the
three high-redshift bins of our sample ($0.45<z<0.75$).
For the photometric sample of counts of neigbors to the WiggleZ markers, 
we count only galaxies with $20 < r'<22.5$ (which is the WiggleZ optical-band
 selection criterion).
For the WiggleZ galaxies, we count all WiggleZ glaxies within the
aperture of each marker, and deem galaxies with a redshift within
$0.002(1+z)$ of a marker galaxy as associated.
We find the ratio of net excess counts to total counts in the aperture 
to be $0.193\pm0.010$ for the photometric sample, where the uncertainty
is based on Poisson statistics.
For the equivalent ratio from the WiggleZ redshift sample, we obtain 
$0.196\pm0.014$ (with 235 out of 1202 galaxies satisfying the redshift
criterion).
Thus, our statistical background subtraction technique produces net
counts that are entirely consistent with results using a sample of
galaxies with known redshifts, further adding confidence to the robustness
of method.

Second, incorrect redshift measurements of the markers will dilute the results.
The redshift plays the key role in computing the distance modulus,  k-correction, and the aperture size for a fixed physical diameter.
The first two impact directly on the rest-frame color and magnitude distributions of the neighbors. 
The latter affects the surface density of the excess counts.
Since the redshift confidence for the markers with redshift quality flag $Q_{zspec}$=3 is $\sim$78.7\%, 
we test the effect of incorrect redshift measurement
 by repeating our analyses using objects with $Q_{zspec}$=4 and 5 only, 
at the cost of having a weaker signal due to stacking fewer color-color-magnitude cubes.
We find that the results are very similar to what we have presented here.

Finally, an interesting question is how often we have another WiggleZ galaxy within the aperture ($r_p$ = 0.25 Mpc) centered at a WiggleZ marker.
We find that $\sim$5-10\% of the markers have at least one other WiggleZ galaxy within a radius of 0.25 Mpc, but the percentage drops to  $\sim$1\% if these other WiggleZ galaxies have to be at a similar redshift ($\Delta z \leq 0.002(1+z)$) as the marker. 
The range in the percentage is patch dependent, as the number density of WiggleZ galaxies varies in different patches.
We also find that the number of the enclosed WiggleZ galaxies is not a strong function of redshift.
Therefore, we conclude that the properties of the WiggleZ neighbors should not be significantly contaminated by the WiggleZ galaxies themselves.

\subsection{The WiggleZ Neighbor Galaxy Luminosity Function}

\subsubsection {The Schechter Function Fit}
Our WiggleZ galaxy neighbor sample represents a complete census of galaxies
in the neighborhood of star-forming galaxies over the redshift range of 0.25 to
0.75. 
On the color-magnitude plane, we separate the galaxy sample into 
red and blue galaxy subsamples.
The GLFs for both the `Blue' and `Red' galaxy samples can be fitted very well 
with single Schechter functions.  Based on the low-redshift bins, where
there is sufficient depth to measure the faint-end slope unambiguously,
they have different shapes such that the red galaxies have a dipping
faint-end (best fitted with $\alpha\gtapr-0.6$), whereas the blue
galaxies have a steep rising faint-end which is best fitted with
 $\alpha\sim-1.3$.  
This is similar in general to what is seen in clusters,
where red-sequence galaxies typically have a turn over in the GLF,
while the blue galaxies increase in number steeply at the faint
end \citep[e.g.,][]{2007ApJ...671.1471B,2009MNRAS.393.1302W}.

  We can make a direct comparison between the WiggleZ neighbor
sample, representing galaxies associated with star-forming galaxies,
and the neighbors of the RCS-WSF sample, representing galaxies in dense 
cluster environments, by combining the $z\sim0.3$ and 0.4 subsamples.
We find the `Blue' galaxy samples to have essentially identical
GLF parameters: 
$\alpha=-1.42\pm0.08$ and $-1.57\pm0.13$, and
$^{0.5}M_r^* = -21.15\pm0.19$ and $-21.37\pm 0.47$. 
The red galaxy GLFs for the two samples appear to have different
faint-end slopes at the $4.3\sigma$ level, 
with $\alpha=-0.18\pm0.17$, and
$-0.74\pm0.06$ 
for the WiggleZ neighbors and the RCS-WSF neighbors.
The RCS-WSF red galaxy neighbors also have a marginally brighter $^{0.5}M_r^*$
of $-20.78\pm0.09$, compared to $-20.42\pm0.15$ for the WiggleZ
neighbor sample, at the $2\sigma$ level.
However, this difference is most likely a reflection of the 
different $\alpha$ used in the best fits;  
fitting both `Red' GLFs using a common $\alpha=-0.5$ for $z=0.25$ to 0.45, 
we obtained 
\M05star=$-20.42\pm0.09$ and $-20.53\pm0.07$
 for the RCS-WSF and the WiggleZ neighbor sample, respectively.
We will further discuss the difference in the GLF shape in \S\ref{subsec:evol}.

\subsubsection {The Evolution of the Galaxy Luminosity Function}

We use the parameter $Q$ (see $\S$\ref{subsec:LF})
to measure the evolution of $M^*$, with the 
asssumption of $\alpha$ being constant with redshift for 
the faint-end of the GLF. 
The results for different samples and $\alpha$'s are shown in Fig. \ref{fig:Q}.
There is a general brightening of $M^*$ with larger redshift.
An intereting trend is that that red-sequence galaxies may have a
lower $Q$ value than that of blue-cloud galaxies, $\sim-1.6$ versus $\sim-2.1$,
but only at the 1.3$\sigma$.

Using a sample drawn from the DEEP2 and COMBO-17 data and with
SDSS data as the local universe epoch,
 \citet{2007ApJ...665..265F}~studied the evolution of the GLF in the
redshift range of  $0.1\leq z<1$ in the $M_B$ band, which is similar in
 rest wavelength to our $^{0.5}M_{r'}$ band.  
They obtained $Q$ values of $-1.23\pm0.36$, $-1.34\pm0.22$, $-1.20\pm0.21$,
over the range of $z\sim0$ to $1$, for their `All', `Blue', and `Red'
samples.
These values appear to be lower from those derived in our samples
(see Table \ref{tab:LF})
at moderately significant levels, especially for the `Blue' sample.
However, the comparison is much more similar if we limit their data
to the same redshift range as ours.
In Figure \ref{fig:Q}, we plot their $M_B^*$ data points.
For a more direct comparison,
we also convert their $B$ band to our $^{0.5}M_{r'}$
by applying color corrections based on
$B-r'$ colors from GISSEL \citep{2003MNRAS.344.1000B} to convert $M_B$ to $M_{r'}$
at zero redshift, and then k-correct $M_{r'}$ to $^{0.5}M_{r'}$.
These data are also plotted in Figure \ref{fig:Q}.

We refit the $Q$ factor from \citet{2007ApJ...665..265F} using their data within the redshift range covered by our the WiggleZ neighbor sample.
We find $Q=$ [$-1.45\pm0.25, -2.02\pm0.32, -0.67\pm0.38$] 
(in their $B$ band) for their
`All', `Blue', and `Red' samples, respectively, 
compared to our values of 
$-1.31\pm0.24$, $-2.10\pm0.26$ ($\alpha$=-1.3) and $-1.59\pm0.30$
 ($\alpha=-0.4$) in Table \ref{tab:LF}.
Thus, the derived $Q$ values from the two studies over the same
redshift range are similar, especially for the `All' and `Blue' samples.
For the `Red' samples, the WiggleZ neighbors have a steeper evolution, 
at the $\sim2\sigma$ level. 
We note that over this redshift range the Faber et al.~results also
produce a relatively significant lower $Q$ value for their `Red' sample 
compared to that for their `Blue' sample, at the 2.7$\sigma$ level,
reinforcing a possible similar trend in the WiggleZ samples. 
The marginally steeper evolution of the `Red' samples in the two
data sets could be a reflection of the difference in the selection 
for the red galaxy samples.
This possible discrepancy could be an indication that there is a difference
in the evolution of early-type galaxies based on the environment.
The red galaxies in the WiggleZ neighbor sample are primarily in
low density regions, conducive to star-formation; while those in the Faber 
et al. sample include red galaxies in all environments, with an expected 
bias towards high density regions.
Thus, it is perhaps not surprising that the two samples have different
$Q$ values, with the red GLF in low-density regions evolving more rapidly.
However, we note that the differences discussed in this subsection are
at the $\sim2\sigma$ levels.  Considerably more detailed studies are
need to firmly establish the differences in the evolution of the GLF
of different galaxy populations in different environments.

\subsection{The Red-Galaxy Fraction}
When the star forming activity in a galaxy ceases, the galaxy ought to become redder in colors.
In Fig. \ref{fig:fred} we explore the redshift dependence of
the red galaxy fraction, $f_{red}$.
Since the dependence of $f_{red}$ on $z$ is similar for samples of different depths, 
for the remainder of the analysis, we use the $^{0.5}M_{r'}$$<$$^{0.5}M_{r'}^*+1.4$ sample. 
which has \fred~measurements with the smallest error bars and is complete for all the redshift bins, unless noted otherwise.

\subsubsection{The Redshift Dependence of \fred}
We observe that $f_{red}$ is remarkably similar at $0.25 \leq z < 0.65$,
with a value of $\sim0.28$.
A linear fit to the 4 points gives a slope of $-0.070\pm0.098$.
The \fred~drops to about 0.20 for the $z=0.7$ bin.
This drop, between the $z=0.6$ and 0.7 bins, is statistically significant
at the $\sim3.5\sigma$ level. 
If we assume the best linear fit for \fred~from the four data points
with $z<0.65$, the \fred~at $z=0.7$
is 4.8$\sigma$ below the extrapolation from the fit.
An alternative simpler description of the trend is  a linear decrease;
however, this linear fit, to all 5 points, has a reduced $\chi^2$ of 3.1, 
indicating that it is not a good description of the data.

To investigate whether there is a change in the dependence of \fred~on $z$
at $z\sim0.7$, we extend the measurement of \fred~to $z\sim0.8$.
The WiggleZ sample has a large number of galaxies at $0.75<z<0.85$,
which we have not included in our analysis because the RCS2 photometry
is only complete to a relatively shallow absolute magnitude limit of --20.6.
Nevertheless, this bin is complete to $^{0.5}M_{r'}$+1.2, and hence, we 
perform the same analysis using the 5819 WiggleZ galaxies in this
redshift bin as markers.
The \fred~computed using this sample is plotted on Figure \ref{fig:fred}.
It shows a continuous drop from the $z=0.7$ data point.
A linear fit applied to all 6 data points from $z=0.3$ to $z=0.8$
with a limiting magnitude of $M^*+1.2$,
produces a reduced $\chi^2$ of 6.3, indicating a poor fit.
Thus, the addition of the higher redshift data
adds confidence to the conclusion there is an on-set of a significant
drop in the red-galaxy fraction at $z>0.65$.

We compare our results to \citet{2010A&A...509A..40I}, who 
derived the blue galaxy fraction $F_{blue}$ for galaxy samples in different 
environments from the zCOSMOS survey.
Of particular interest is their `All' and `Isolated' subsamples of
Sample III, with a depth of $M=M^*+1.5$, covering the redshift
range of $\sim0.2$ to 0.6.  \citet{2010A&A...509A..40I} 
parametrize the evolution of $F_{blue}$ by a power law
of the form equivalent to: \fred$(z)=1-F_{blue}(0)(1+z)^\beta$, with
$F_{blue}(0)=0.58\pm 0.10$ and $\beta=0.69\pm 0.44$ for the Isolated
sample, using 3 data points.
We plot their $F_{blue}(z)$ as \fred~on Figure
\ref{fig:fred} using their parametrization.
Our \fred~has values more similar to their Isolated sample, and well
below their `All' sample.
This is consistent with our expectation that the WiggleZ neighborhoods
have by-and-large a low galaxy density environment, dominated by
star-forming galaxies.

For a more direct comparison, we fit our \fred~with the power law 
over a longer redshift range ($z=0.20-0.70$) than \citet{2010A&A...509A..40I}, and obtain (for $M^*+1.4$):
$F_{blue}(0)=0.59\pm0.04$, and $\beta=0.52\pm0.13$ with a reduced $\chi^2\sim3.1$.
Thus, our data show a similar, rate of
decrease of \fred~with increasing redshift as that found by \citet{2010A&A...509A..40I}.
However, 
we note that both of the fitting models to our data--the linear fit and 
the power-law fit--have reduced $\chi^2$s considerably larger than 1:
$\sim3-4$ for both using data up to $z=0.7$ and $z=0.8$.
This indicates that a simple continuous decrease of \fred~with
redshift is likely not a good description of its evolution.

Thus, an interesting result is that our data, covering a longer
redshift baseline than the \citet{2010A&A...509A..40I}~study, show
a more or less constant \fred~up to $z\sim0.6$ before seeing a drop.
Such a description of the change in \fred~for galaxies in poor environments
with redshift is in fact also consistent with the data of the `Isolated' 
sample of \citet{2010A&A...509A..40I}, with their three $F_{blue}$ data 
points covering the redshift between 0.2 and 0.6 being consistent with having
similar values within their uncertainties.
We note that the more or less constant \fred$\sim0.3$ for $z\ltapr0.6$
for the WiggleZ neighbor samples is similar to that obtained for
galaxies in low-density regions at the local universe at $z\sim0$;
e.g.,
 \citet{2004ApJ...615L.101B} derived $f_{red}\sim0.35$ from the SDSS sample 
for their low local-galaxy density samples of galaxies of $M_V\ltapr-20$.
The \fred~for the WiggleZ galaxy neighbors extrapolate to $\sim0.32$ at $z=0$.
Thus, there appears to be a relatively small amount of evolution
in \fred~in low
galaxy density regions from $z\sim0.6$ all the way to $z\sim0$.

Because the WiggleZ sample is selected in part by observed UV and optical 
fluxes of the objects, this abrupt increase in the change 
in \fred~between the 0.6 and 0.7 redshift
bins could conceivably be contributed by some unknown correlation or
thresholding effect between the star-formation rate or luminosity of
the markers and the properties of their neighbors.
To test the effect of the UV flux selection criterion, we select two
subsamples of markers within an identical UV absolute luminosity range which is
complete in the both $z\sim0.6$ and $z\sim0.7$ bins ($M_{NUV}\leq -20.7$), and compute their \fred.
We obtain \fred=0.293$\pm$0.022, and 0.207$\pm$0.017  for the 
neighbors in the two 
redshift bins, indicating a similar and significant (at $\sim3\sigma$)
drop to that when the whole sample without control on $M_{NUV}$  is used.
Similarly, we compare the \fred~of neighbors of subsamples of markers with the 
same $r'$ absolute magnitude ($^{0.5}M_{r'}$ between --21.5 and --22.5) 
in the last two redshift bins,
and obtain  \fred=0.255$\pm$0.045 and 0.176$\pm$0.033; again, showing
a drop similar to that obtained using the whole redshift bins.
Thus, we can conclude that the drop in \fred~between $z\sim0.6$
and 0.7 is likely not a result of the different star-formation properties of
the central markers.

\subsubsection{The Evolution of the Red-Galaxy Fraction} \label{subsubsec:fred_evol}

\begin{figure}
\includegraphics[scale=0.3]{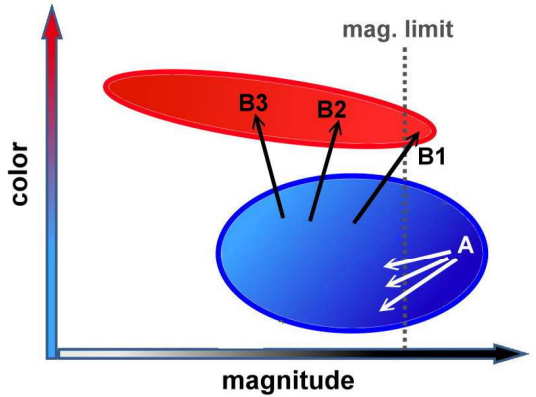}
\caption{ Schematic color-magnitude diagram showing the flows of galaxies
into and out of the blue cloud.
The narrow red ellipse represents the location of the red sequence;
the blue ellipse marks the blue cloud.
The vertical dotted line indicates the magnitude limit of the sample
for computing \fred.
The white arrows (A) show the replenishment of galaxies in the blue
cloud due to an increase in the star-formation rate of blue galaxies
fainter than the magnitude limit.
The black three arrows illustrate the possible flows of galaxies from the
blue cloud into the red-sequence after star formation quenched.
Arrow B3 represents star-formation being quenched by an approximately
equal-mass merger.
Arrow B2 stands for the quenching of star formation, 
 while arrow B1 represents the case when the galaxy moves back beyond the magnitude limit again due to its quenched star formation.
  \label{fig:cmd_schematic}}
\end{figure}

Our data allow us to look at the redshift evolution of the red-galaxy
fraction \fred~up to $z\sim0.7$ using samples complete to an 
absolute luminosity about 1.5 mag beyond $M^*$.
Under the simplest assumption of a sample of galaxies in a closed
volume, \fred~allows us to follow the end result
of the process of galaxies having their star formation quenched and
eventually turning red, becoming a member of the red sequence \citep[e.g.,][]{2001MNRAS.321...18K}.
However, the use of a luminosity limit produces ambiguities in how to 
interpret the average change in the galaxy population between the different
epochs, since galaxies of a given mass may enter and leave the
sample depending on their star-formation state.
To assist in the interpretation of the observed change in \fred~with
redshift, we illustrate in Figure \ref{fig:cmd_schematic}
the possible flows of galaxies into and out of luminosity-limited 
blue and red galaxy samples.

 In a simple picture, the number of galaxies in the blue cloud
can be augmented by new star formation in galaxies 
fainter than the luminosity limit, boosting them into the sample,
adding to the blue galaxy counts.  This replenishment is represented
by the white arrows (A) in Figure \ref{fig:cmd_schematic}.
In a field situation, where infall in general is not a major process,
one can consider this replenishment generally being controlled by the
canonical down-sizing scenario of star formation -- lower mass galaxies
may start forming stars, and in some instances, boosting the luminosity
of the galaxy into the blue cloud sample.

  On the other hand, one would expect a continuous flow of galaxies
from the blue cloud into the red sequence as galaxies evolve.
This would primarily be created by the quenching of star formation in
galaxies in the blue cloud by various processes.
There are several paths that this transformation into the red sequence
may take, and the are illustrated by the black three arrows (B1, B2, and B3)
schematically.
The luminosity of a galaxy will generally fade as its star formation is
quenched and turns red (B1 and B2), and some of them will become fainter than the sample magnitude limit 
and leave the sample.
There is also a possibility that the galaxy becomes brighter
if the cessation of star formation follows a major merger (arrow B3).
However, in poor environments, it is unlikely that this is a dominant
event; e.g., \citet{2008ApJ...683...33H}
found approximately 6\%~of 
galaxies may have undergone a major merger since $z\sim0.8$.

  In the WiggleZ sample of neighbors of strong star-forming galaxies we
find that \fred~is essentially flat from $z\sim0.6$ to the present, with
evidence of a decrease at $z\gtapr0.6$.
In the simplest picture, the change in \fred~over redshift can be
interpreted as a change in the relative flow rates of galaxies
into the red sequence and into the blue cloud.
Thus, at $z\ltapr0.6$ we can effectively conclude that the rate of
 transformation of blue galaxies into red galaxies, which builds up the
red-sequence, is approximately equal to the replenishment rate of
blue galaxies brighter than the sampling limit; while at $z\gtapr0.6$
the rate of transformation of blue galaxies into red galaxies 
is lower than the replenishment rate of blue galaxies, so that there
is a net decrease of red galaxies relative to the blue galaxies.

However, the change in \fred~with redshift does not give us
unambiguous information on the actual changes in these rates.
If we assume that the rate of the build-up of the red sequence is constant
over the redshift range of $\sim0.8$ to 0.2, we would then conclude
that the rate of replenishment of blue galaxies (into the luminosity
limited sample) is decreasing from $z\sim0.8$ to 0.6, and then
becomes stable, or decreasing at a much lower rate, at $z\ltapr0.6$.
Alternatively, if we assume that the replenishment of blue galaxies
is constant, then the rate of transformation of blue cloud galaxies
into the red sequence is increasing from $z\sim0.8$ to 0.6, and
become stable at $z\ltapr0.6$.
The detailed picture is certain to be more complex.
In a more general picture, the environment will be a major factor that
affects these rates of flows.
An important effect to consider is the infall of galaxies into a high density
region such as the parent halo of a galaxy cluster or group.
We will discuss this effect in \S\ref{subsec:evol}, in conjunction
with the evolution of \fred~in clusters.


\subsubsection{The \fred~of the Cluster Neighborhood Sample}
We  have also derived \fred~for the RCS-WSF cluster neighborhood sample for the
$z\sim0.3$ and $z\sim0.4$ bins, which are plotted as open circles in Fig. \ref{fig:fred}.
They show a clear difference due to the environments of the galaxies;
the cluster neighborhood samples have \fred~of $\sim0.8$, indicative
of high galaxy density regions.
While we do not have the redshift range to examine the evolution of
\fred~in clusters, studies using similar techniques of large samples
of clusters show a continuous decrease of \fred~with redshift, i.e., the
Butcher-Oemler effect (Butcher \& Oemler, 1984).
For instances, Loh et al. (2008), using a sample of approximately 1000 clusters
 from RCS1 at $0.4\leq z \leq 0.9$, show a steady decrease in \fred~of
about 0.4; while the spectroscopic sample of \citet{2001ApJ...547..609E}
indicates a steady drop of \fred~from 0.9 to 0.7 at $z$=0.2-0.5.

Comparing these results with the WiggleZ neighbor sample,
there appears to be a rather different behavior in the change in \fred~as
a function of redshift for galaxies in regions around
star-forming galaxies from those in high galaxy density regions
around massive halos.
Instead of a steady decline in \fred~with increasing redshift, the \fred~for
the WiggleZ neighbors have a basically flat dependence on $z$, and
show a significant drop only at $z\gtapr0.65$.
Similarly, Iovino et al.~(2010) show that galaxies classified as being
in a group environment have a much steeper dependence of their blue-galaxy
fraction on redshift than those considered to be in isolated environments.
This difference in the evolution of \fred~can be considered as a clear 
demonstration of the effect of environment on the evolution timescale 
of galaxy populations.

\subsection{Galaxy Evolution and Environment} \label{subsec:evol}
The topic of galaxy evolution has been studied in galaxy clusters over many decades. 
With modern large surveys, the focus has extended to field galaxies. 
Here, we would like to use the term `field' to refer to all galaxies in a 
general blind field survey, such as SDSS or 2dF, where galaxy density ranges 
from that of isolated galaxies to clusters.
Studies using such `field' samples have revealed a strong correlation between many properties of galaxies and environment (usually parametrized by local galaxy density). 
One example is that the fraction of star-forming (or passively evolving
galaxies) changes strongly with local galaxy density \citep[e.g.,][]{2006MNRAS.373..469B,2009ApJ...698...83L,2010A&A...509A..40I}.
Other works focus on star-forming galaxies. 
These star-forming galaxies naturally reside in regions beyond galaxy clusters.
Their star-formation rates and colors are found not to have
 a strong dependence on environment \citep[e.g.,][]{2001ApJ...559..606C,2005AJ....130.1482R,2009MNRAS.398..754B,2007ApJS..172..270C}.
Such conclusions appear to be different from the results from using all 
`field' galaxies, and this is likely a reflection of the sample properties.

Our WiggleZ galaxies are members of these star-forming galaxies, selected
by their  detectable UV flux. 
We use the term `WiggleZ neighborhood' to refer to local regions around the WiggleZ galaxies to distinguish it from the general `field' environment. 
 The WiggleZ neighborhood is 
likely low-density environment regions in the large-scale structure.
The galaxy sample generated by counting excess galaxies in the WiggleZ
galaxy neighborhood represents a census of galaxies in 
regions around star-forming galaxies covering a significant redshift range.

We also examine \fred~in dense environments using the neighbors of the
RCS-WSF sample of cluster galaxies.  
The much higher \fred~value ($\sim0.8$ vs $\sim0.3$) measured is 
an indication of the more rapid build-up of the red sequence in dense
environments.
Furthermore, it appears that the evolution of \fred~for these two samples
is also very different, with the WiggleZ neighbors having very moderate
or no increase in \fred~since $z\sim0.6$.

In the context of the discussion of Figure \ref{fig:cmd_schematic} in \S
\ref{subsubsec:fred_evol}, the flow paths and rates of blue and red
galaxies on the CMD in cluster environments would be very different 
from those in the WiggleZ neighborhoods.
Here, the major source of replenishment of blue galaxies in massive dense halos
is likely to be the infall of galaxies along the large-scale structure.
These galaxies are then transformed into red galaxies within some time
scale due to processes such as tidal striping, ram pressure, galaxy
harassment and interactions, which are generally associated with
the quenching of star formation in galaxy groups and clusters.
The continuous increase in \fred~can also be in part attributed
to the cosmological decrease in infall rate with time in a low $\Omega_m$
universe \citep[e.g.,][]{2001ApJ...547..609E}.


There is an additional difference between these two samples in the
shape of the GLF of the red galaxy population.
The Schechter function fit indicates that the red galaxies
in dense environments have a GLF with a steeper faint-end.
This can also be a result of the red sequence in cluster environments
being in a more advanced build-up stage, or, the build-up of the
faint-end of the red sequence in dense environment 
is more rapid than that in regions around star-forming galaxies.
The build-up of the faint-end of the red sequence in clusters
has been traced by a number of studies  
\citep[e.g.,][]{2004ApJ...608..752B,2005MNRAS.362..268T, 2006ApJ...647..853W, 2007ApJ...661...95S, 2009MNRAS.400...68D}.
\citet{2008ApJ...673..742G} show some preliminary evidence that 
this build-up is faster in rich clusters than in poor clusters, which is
consistent with the comparison here between clusters and low galaxy
density regions.
Another apparent difference in the GLF of the red galaxies in the two
sample is that the red galaxies in the cluster sample show a significant
excess to the Schechter function fit at bright magnitudes, suggesting
that there are relatively more massive red galaxies in dense environments, 
either due to initial galaxy formation history or an increased rate of
mergers in the evolution of these galaxies.

The possible difference in the evolution of $M^*$ of the red galaxy
GLFs described in \S5.2.2, while not of high statistical significance,
also fits in with the scenario of the dependence of the evolution 
of the red galaxy population with environment.
The lower value of $Q$ (i.e., slower evolution of $M^*$) for the
Faber et al. (2007) `Red' galaxy sample, which are in an environment
denser than those from the WiggleZ sample, suggests that red
galaxies in denser environments are likely older.
This would be expected if galaxies in dense environments such as
clusters and group turn red and dead earlier in the
history of the universe than those in low-density region.

In summary, in the general scenario of galaxy evolution and its connection to
the environment the WiggleZ neighborhoods can be considered as
regions where there are minimal major environmental events affecting
the evolution of galaxies.
These regions are likely the low galaxy density parts of the
large cosmic structure where star formation is still occurring well
into the current epoch. 
Here, galaxies are unlikely to be affected by environmental influences
that are associated with infall into a massive halo such as a
substantial galaxy group or cluster.
The galaxies in these neighborhoods can be considered as primarily following
a secular galaxy evolution path, controlled by their
own nature at birth, with environmental effects playing a role over
a much longer time scale.
In these environments, only a small fraction ($\sim$20-30\%) of the galaxies have turned red over the redshift range of 0.7 to 0.3.
In a cluster region, where there is a continuous infall of galaxies
from the lower-density environments replenishing the blue cloud, 
the effects of the large dark matter parent halo on galaxies would produce a 
very different mix of galaxy populations.
In these regions, environment plays a dominant role, accelerating the
quenching of star formation, and  transforming
the infallen galaxies into the red sequence over a relatively short
time scale, producing a red sequence in a more advanced evolutionary stage.

\section{Summary} \label{sec:summary}
We have probed galaxy evolution at $0.25 \leq z \leq 0.75$ using optical data from the RCS2 around $\sim$41,000 star-forming spectroscopic galaxies from the WiggleZ project.
Because of the complicated selection criteria in the WiggleZ survey, 
galaxies in the spectroscopic sample have discrete characteristics as a 
function of redshift.
We therefore examine optical properties of galaxies within 0.25 Mpc to WiggleZ galaxies using stacked color-color-magnitude cubes.
The idea is to use the WiggleZ galaxies as markers, and assume that they and the surrounding neighbors are at the same redshift.
By applying background subtraction and stacking the net excess counts
 over a large number of markers, we are able to study 
the properties of the neighbors around the WiggleZ galaxies.

We also examine how the optical colors of the neighbors correlate 
with the WiggleZ [OII]$\lambda$3727\AA~equivalent width, $NUV$ flux, 
and AGN activity indicators 
at different redshifts (Figures \ref{fig:ccplot_O2}, \ref{fig:ccplot_nuv}, and \ref{fig:r1}).
We find in general that the neighbor galaxies populate the same color-color
 spaces without significant dependence on these properties of the markers,
suggesting that the properties of the neighbor galaxies are not 
strongly affected by, or correlate with, the characteristics of the WiggleZ galaxies themselves,
Thus, they can be used to study the evolution of the photometric properties
of galaxies in low-density, star-forming 
regions over the redshift range of 0.25 to 0.75.

Our major findings are: 
\\
\noindent
(1) {\it The majority of WiggleZ neighbors are blue galaxies which have
a steeper faint-end slope and a faster evolution term in $^{0.5}M^*_{r'}$
than the red galaxies}.  

  The CMD of the WiggleZ neighbors shows the characteristic bimodal 
distribution of a red sequence and a blue cloud, with the latter dominating
the galaxy population, containing 65 to 85\% of the galaxies (depending
on the depth of sampling and redshift).  The GLFs of the two populations
can be fitted with single Schechter functions, with the blue galaxies
having a much steeper ($\alpha\sim-1.3$) faint-end than the red galaxies
($\alpha\sim-0.4$).
Based on the three low-redshift bins ($0.25<z<0.55$), where the data are complete to
$^{0.5}M_{r'}$=--19.0, we find no significant changes in the faint-end slope of
the GLFs with redshift.
There is significant evolution in $M^*$ for both the `Blue' and `Red'
subsamples.
We find that blue galaxies have a marginally more rapid evolution in $M^*$
over this redshift range with $Q\sim-2.10$, compared to $Q\sim-1.59$
for the red galaxies.
While the $Q$ values for both the red and blue galaxy samples appear 
to be  steeper than the typical values in the literature,
these evolution factors are similar to those from \citet{2007ApJ...665..265F}
when compared over the same redshift range.
\\

\noindent
(2) {\it The red galaxy fraction $f_{red}$ in the WiggleZ neighborhood is 
approximately constant since $z\sim0.6$, but drops at $z\gtapr0.7$}. 

The WiggleZ neighbor galaxies have a red-galaxy fraction (\fred)
considerably smaller than that of the neighbors of the RCS-WSF sample of 
markers, at $0.25<z<0.45$, obtained the same way as the WiggleZ sample.
The evolution of the WiggleZ neighbor \fred~with redshift is modest; 
\fred~can be described as basically flat as a function of redshift with only a very moderate decrease up to $z\sim0.6$.
The average \fred~value over the redshift range of $0.25<z<0.65$ is
$\sim0.30$, similar to field galaxies at $z\sim0$.
A large drop to \fred$\sim0.20$ is seen for the $z\sim0.7$ sample.
This drop is confirmed by extending the measurement of \fred~to $z\sim0.8$
using a sample with a slightly brighter absolute magnitude limit.
Furthermore, this drop does not seem to be associated with the
larger average luminosity of the markers at the higher $z$ bin.
The change in $f_{red}$ with redshift in the WiggleZ neighborhood can be seen as either a higher rate of relatively bright ($M\ltapr M^*+1$) star-forming galaxies entering the luminosity limited sample at $z\sim$0.7, 
or a decrease in the quenching rate of star formation at this redshift.
\\

\noindent
(3) {\it The comparison between the WiggleZ and RCS-WSF neighbor samples shows an environmental influence on galaxy properties and evolution, with the
red sequence in cluster environment being in a more advanced build-up stage.} 

We examine the effects of environment on the galaxy population properties
by comparing the WiggleZ neighborhood galaxies to that of the RCS-WSF 
neighbor sample.
Besides the expected and obvious difference in \fred~values of the two samples
(with \fred, at $\sim0.8$, being much larger in the RCS-WSF sample), we also find
significant difference in the GLFs of the red galaxies of the two samples.
The faint-end slope $\alpha$ for the red GLF
of the RCS-WSF sample is considerably steeper ($\sim-0.7$ vs $\sim-0.4$).
This can be taken as the build-up of the faint-end of the red sequence
being in a more advanced stage in rich environments (such galaxy clusters
and groups) than that in lower galaxy density regions around 
star-forming galaxies. Furthermore, there is also evidence that there
are excess luminous red galaxies in the RCS-WSF sample.
These findings point to the importance of environment in affecting the
history of star formation in galaxies. 
Galaxies in cluster/group environments likely have suffered significant environmental events that rapidly shut down their star formation, turning the galaxy red.
Whereas in regions where star formation is still prevalent, environmental events likely occur much less frequently and their effects spread over a longer time
scale, delaying the build-up of the red sequence.

\acknowledgements
The RCS2 data in this paper are 
based on observations obtained with MegaPrime/MegaCam, a joint project of CFHT and CEA/DAPNIA, at the Canada-France-Hawaii Telescope (CFHT) which is operated by the National Research Council (NRC) of Canada, the Institute National des Sciences de l'Univers of the Centre National de la Recherche Scientifique of France, and the University of Hawaii.
I.H.L. wishes to thank to the Australian Research Council Linkage International Grant for the early development of this work. I.H.L. and H.K.C.Y. thank the Academia Sinica Institute of Astronomy and Astrophysics, Taiwan, for their hospitality during the early stage of the writing of the paper.
The RCS and the research of H.K.C.Y. are supported by grants from the Natural Science and Engineering Research Council of Canada and the Canada Research Chair program.
The WiggleZ team acknowledges financial support from the Australian Research Council through Discovery Project grants. 
The WiggleZ survey would not be possible without the dedicated work of the staff of the Anglo-Australian Observatory in the development and support of the AAOmega spectrograph, and the running of the AAT.



\begin{thebibliography}{74}
\expandafter\ifx\csname natexlab\endcsname\relax\def\natexlab#1{#1}\fi

\bibitem[{{Adelman-McCarthy et al.}(2006)}]{2006ApJS..162...38A}
{Adelman-McCarthy et al.}, J.~K. 2006, \apjs, 162, 38

\bibitem[{{Baldry} {et~al.}(2006){Baldry}, {Balogh}, {Bower}, {Glazebrook},
  {Nichol}, {Bamford}, \& {Budavari}}]{2006MNRAS.373..469B}
{Baldry}, I.~K., {Balogh}, M.~L., {Bower}, R.~G., {Glazebrook}, K., {Nichol},
  R.~C., {Bamford}, S.~P., \& {Budavari}, T. 2006, \mnras, 373, 469

\bibitem[{{Baldwin} {et~al.}(1981){Baldwin}, {Phillips}, \&
  {Terlevich}}]{1981PASP...93....5B}
{Baldwin}, J.~A., {Phillips}, M.~M., \& {Terlevich}, R. 1981, \pasp, 93, 5

\bibitem[{{Balogh} {et~al.}(2004){Balogh}, {Baldry}, {Nichol}, {Miller},
  {Bower}, \& {Glazebrook}}]{2004ApJ...615L.101B}
{Balogh}, M.~L., {Baldry}, I.~K., {Nichol}, R., {Miller}, C., {Bower}, R., \&
  {Glazebrook}, K. 2004, \apjl, 615, L101

\bibitem[{{Balogh} {et~al.}(1997){Balogh}, {Morris}, {Yee}, {Carlberg}, \&
  {Ellingson}}]{1997ApJ...488L..75B}
{Balogh}, M.~L., {Morris}, S.~L., {Yee}, H.~K.~C., {Carlberg}, R.~G., \&
  {Ellingson}, E. 1997, \apjl, 488, L75+

\bibitem[{{Balogh et al.}(2009)}]{2009MNRAS.398..754B}
{Balogh et al.}, M.~L. 2009, \mnras, 398, 754

\bibitem[{{Barkhouse} {et~al.}(2007){Barkhouse}, {Yee}, \&
  {L{\'o}pez-Cruz}}]{2007ApJ...671.1471B}
{Barkhouse}, W.~A., {Yee}, H.~K.~C., \& {L{\'o}pez-Cruz}, O. 2007, \apj, 671,
  1471

\bibitem[{{Barnes} \& {Hernquist}(1991)}]{1991ApJ...370L..65B}
{Barnes}, J.~E. \& {Hernquist}, L.~E. 1991, \apjl, 370, L65

\bibitem[{{Bell et al.}(2004)}]{2004ApJ...608..752B}
{Bell et al.}, E.~F. 2004, \apj, 608, 752

\bibitem[{{Bertelli} {et~al.}(1994){Bertelli}, {Bressan}, {Chiosi}, {Fagotto},
  \& {Nasi}}]{1994A&AS..106..275B}
{Bertelli}, G., {Bressan}, A., {Chiosi}, C., {Fagotto}, F., \& {Nasi}, E. 1994,
  \aaps, 106, 275

\bibitem[{{Blake et al.}(2011)}]{2011MNRAS.415.2892B}
{Blake}, C., 2011, \mnras, 415, 2892

\bibitem[{{Blanton}(2006)}]{2006ApJ...648..268B}
{Blanton}, M.~R. 2006, \apj, 648, 268

\bibitem[{{Blanton et al.}(2003)}]{2003ApJ...594..186B}
{Blanton et al.}, M.~R. 2003, \apj, 594, 186

\bibitem[{{Bongiorno et al.}(2010)}]{2010A&A...510A..56B}
{Bongiorno et al.}, A. 2010, \aap, 510, A56+

\bibitem[{{Bruzual} \& {Charlot}(2003)}]{2003MNRAS.344.1000B}
{Bruzual}, G. \& {Charlot}, S. 2003, \mnras, 344, 1000

\bibitem[{{Butcher} \& {Oemler}(1984)}]{1984ApJ...285..426B}
{Butcher}, H. \& {Oemler}, Jr., A. 1984, \apj, 285, 426

\bibitem[{{Cannon et al.}(2006)}]{2006MNRAS.372..425C}
{Cannon et al.}, R. 2006, \mnras, 372, 425

\bibitem[{{Carter} {et~al.}(2001){Carter}, {Fabricant}, {Geller}, {Kurtz}, \&
  {McLean}}]{2001ApJ...559..606C}
{Carter}, B.~J., {Fabricant}, D.~G., {Geller}, M.~J., {Kurtz}, M.~J., \&
  {McLean}, B. 2001, \apj, 559, 606

\bibitem[{{Cassata et al.}(2007)}]{2007ApJS..172..270C}
{Cassata et al.}, P. 2007, \apjs, 172, 270

\bibitem[{{Chabrier}(2003)}]{2003PASP..115..763C}
{Chabrier}, G. 2003, \pasp, 115, 763

\bibitem[{{Christlein} {et~al.}(2009){Christlein}, {Gawiser}, {Marchesini}, \&
  {Padilla}}]{2009MNRAS.400..429C}
{Christlein}, D., {Gawiser}, E., {Marchesini}, D., \& {Padilla}, N. 2009,
  \mnras, 400, 429

\bibitem[{{Colless et al.}(2001)}]{2001MNRAS.328.1039C}
{Colless et al.}, M. 2001, \mnras, 328, 1039

\bibitem[{{Cooper et al.}(2007)}]{2007MNRAS.376.1445C}
{Cooper et al.}, M.~C. 2007, \mnras, 376, 1445

\bibitem[{{De Lucia} {et~al.}(2009){De Lucia}, {Poggianti}, {Halliday},
  {Milvang-Jensen}, {Noll}, {Smail}, \& {Zaritsky}}]{2009MNRAS.400...68D}
{De Lucia}, G., {Poggianti}, B.~M., {Halliday}, C., {Milvang-Jensen}, B.,
  {Noll}, S., {Smail}, I., \& {Zaritsky}, D. 2009, \mnras, 400, 68

\bibitem[{{De Robertis} {et~al.}(1998){De Robertis}, {Yee}, \&
  {Hayhoe}}]{1998ApJ...496...93D}
{De Robertis}, M.~M., {Yee}, H.~K.~C., \& {Hayhoe}, K. 1998, \apj, 496, 93

\bibitem[{{Dressler} \& {Gunn}(1983)}]{1983ApJ...270....7D}
{Dressler}, A. \& {Gunn}, J.~E. 1983, \apj, 270, 7

\bibitem[{{Dressler et al.}(1980)}]{1980ApJ...236..351D}
{Dressler et al.}, A. 1980, \apj, 236, 351

\bibitem[{{Drinkwater et al.}(2010)}]{2010MNRAS.401.1429D}
{Drinkwater et al.}, M.~J. 2010, \mnras, 401, 1429

\bibitem[{{Ellingson} {et~al.}(2001){Ellingson}, {Lin}, {Yee}, \&
  {Carlberg}}]{2001ApJ...547..609E}
{Ellingson}, E., {Lin}, H., {Yee}, H.~K.~C., \& {Carlberg}, R.~G. 2001, \apj,
  547, 609

\bibitem[{{Faber et al.}(2007)}]{2007ApJ...665..265F}
{Faber et al.}, S.~M. 2007, \apj, 665, 265

\bibitem[{{Gehrels}(1986)}]{1986ApJ...303..336G}
{Gehrels}, N. 1986, \apj, 303, 336

\bibitem[{{Gilbank} {et~al.}(2011){Gilbank}, {Gladders}, {Yee}, \&
  {Hsieh}}]{2011AJ....141...94G}
{Gilbank}, D.~G., {Gladders}, M.~D., {Yee}, H.~K.~C., \& {Hsieh}, B.~C. 2011,
  \aj, 141, 94

\bibitem[{{Gilbank} {et~al.}(2008){Gilbank}, {Yee}, {Ellingson}, {Gladders},
  {Loh}, {Barrientos}, \& {Barkhouse}}]{2008ApJ...673..742G}
{Gilbank}, D.~G., {Yee}, H.~K.~C., {Ellingson}, E., {Gladders}, M.~D., {Loh},
  Y., {Barrientos}, L.~F., \& {Barkhouse}, W.~A. 2008, \apj, 673, 742

\bibitem[{{Gilbank et al.}(2010)}]{2010MNRAS.tmp..704G}
{Gilbank et al.}, D.~G. 2010, \mnras, 704

\bibitem[{{Gladders} {et~al.}(1998){Gladders}, {Lopez-Cruz}, {Yee}, \&
  {Kodama}}]{1998ApJ...501..571G}
{Gladders}, M.~D., {Lopez-Cruz}, O., {Yee}, H.~K.~C., \& {Kodama}, T. 1998,
  \apj, 501, 571

\bibitem[{{Haines et al.}(2009)}]{2009ApJ...704..126H}
{Haines et al.}, C.~P. 2009, \apj, 704, 126

\bibitem[{{Hopkins} \& {Beacom}(2006)}]{2006ApJ...651..142H}
{Hopkins}, A.~M. \& {Beacom}, J.~F. 2006, \apj, 651, 142

\bibitem[{{Hsieh} {et~al.}(2008){Hsieh}, {Yee}, {Lin}, {Gladders}, \&
  {Gilbank}}]{2008ApJ...683...33H}
{Hsieh}, B.~C., {Yee}, H.~K.~C., {Lin}, H., {Gladders}, M.~D., \& {Gilbank},
  D.~G. 2008, \apj, 683, 33

\bibitem[{{Iovino et al.}(2010)}]{2010A&A...509A..40I}
{Iovino et al.}, A. 2010, \aap, 509, A40+

\bibitem[{{Kaviraj} {et~al.}(2005){Kaviraj}, {Devriendt}, {Ferreras}, \&
  {Yi}}]{2005MNRAS.360...60K}
{Kaviraj}, S., {Devriendt}, J.~E.~G., {Ferreras}, I., \& {Yi}, S.~K. 2005,
  \mnras, 360, 60

\bibitem[{{Kodama} {et~al.}(1998){Kodama}, {Arimoto}, {Barger}, \&
  {Arag'on-Salamanca}}]{1998A&A...334...99K}
{Kodama}, T., {Arimoto}, N., {Barger}, A.~J., \& {Arag'on-Salamanca}, A. 1998,
  \aap, 334, 99

\bibitem[{{Kodama} \& {Bower}(2001)}]{2001MNRAS.321...18K}
{Kodama}, T. \& {Bower}, R.~G. 2001, \mnras, 321, 18

\bibitem[{{Lamareille} {et~al.}(2004){Lamareille}, {Mouhcine}, {Contini},
  {Lewis}, \& {Maddox}}]{2004MNRAS.350..396L}
{Lamareille}, F., {Mouhcine}, M., {Contini}, T., {Lewis}, I., \& {Maddox}, S.
  2004, \mnras, 350, 396

\bibitem[{{Lemaux} {et~al.}(2010){Lemaux}, {Lubin}, {Shapley}, {Kocevski},
  {Gal}, \& {Squires}}]{2010ApJ...716..970L}
{Lemaux}, B.~C., {Lubin}, L.~M., {Shapley}, A., {Kocevski}, D., {Gal}, R.~R.,
  \& {Squires}, G.~K. 2010, \apj, 716, 970

\bibitem[{{Li} {et~al.}(2009){Li}, {Yee}, \& {Ellingson}}]{2009ApJ...698...83L}
{Li}, I.~H., {Yee}, H.~K.~C., \& {Ellingson}, E. 2009, \apj, 698, 83

\bibitem[{{Lilly et al.}(2007)}]{2007ApJS..172...70L}
{Lilly et al.}, S.~J. 2007, \apjs, 172, 70

\bibitem[{{Lin} {et~al.}(1999){Lin}, {Yee}, {Carlberg}, {Morris}, {Sawicki},
  {Patton}, {Wirth}, \& {Shepherd}}]{1999ApJ...518..533L}
{Lin}, H., {Yee}, H.~K.~C., {Carlberg}, R.~G., {Morris}, S.~L., {Sawicki}, M.,
  {Patton}, D.~R., {Wirth}, G., \& {Shepherd}, C.~W. 1999, \apj, 518, 533

\bibitem[{{Liu} {et~al.}(2008){Liu}, {Capak}, {Mobasher}, {Paglione}, {Rich},
  {Scoville}, {Tribiano}, \& {Tyson}}]{2008ApJ...672..198L}
{Liu}, C.~T., {Capak}, P., {Mobasher}, B., {Paglione}, T.~A.~D., {Rich}, R.~M.,
  {Scoville}, N.~Z., {Tribiano}, S.~M., \& {Tyson}, N.~D. 2008, \apj, 672, 198

\bibitem[{{Loh} {et~al.}(2008){Loh}, {Ellingson}, {Yee}, {Gilbank}, {Gladders},
  \& {Barrientos}}]{2008ApJ...680..214L}
{Loh}, Y., {Ellingson}, E., {Yee}, H.~K.~C., {Gilbank}, D.~G., {Gladders},
  M.~D., \& {Barrientos}, L.~F. 2008, \apj, 680, 214

\bibitem[{{Madau} {et~al.}(1998){Madau}, {Pozzetti}, \&
  {Dickinson}}]{1998ApJ...498..106M}
{Madau}, P., {Pozzetti}, L., \& {Dickinson}, M. 1998, \apj, 498, 106

\bibitem[{{Mahajan} \& {Raychaudhury}(2009)}]{2009MNRAS.400..687M}
{Mahajan}, S. \& {Raychaudhury}, S. 2009, \mnras, 400, 687

\bibitem[{{Martin et al.}(2005)}]{2005ApJ...619L...1M}
{Martin et al.}, D.~C. 2005, \apjl, 619, L1

\bibitem[{{Moore} {et~al.}(1996){Moore}, {Katz}, {Lake}, {Dressler}, \&
  {Oemler}}]{1996Natur.379..613M}
{Moore}, B., {Katz}, N., {Lake}, G., {Dressler}, A., \& {Oemler}, A. 1996,
  \nat, 379, 613

\bibitem[{{Rines} {et~al.}(2005){Rines}, {Geller}, {Kurtz}, \&
  {Diaferio}}]{2005AJ....130.1482R}
{Rines}, K., {Geller}, M.~J., {Kurtz}, M.~J., \& {Diaferio}, A. 2005, \aj, 130,
  1482

\bibitem[{{Rola} {et~al.}(1997){Rola}, {Terlevich}, \&
  {Terlevich}}]{1997MNRAS.289..419R}
{Rola}, C.~S., {Terlevich}, E., \& {Terlevich}, R.~J. 1997, \mnras, 289, 419

\bibitem[{{Salimbeni} {et~al.}(2008){Salimbeni}, {Giallongo}, {Menci},
  {Castellano}, {Fontana}, {Grazian}, {Pentericci}, {Trevese}, {Cristiani},
  {Nonino}, \& {Vanzella}}]{2008A&A...477..763S}
{Salimbeni}, S., {Giallongo}, E., {Menci}, N., {Castellano}, M., {Fontana}, A.,
  {Grazian}, A., {Pentericci}, L., {Trevese}, D., {Cristiani}, S., {Nonino},
  M., \& {Vanzella}, E. 2008, \aap, 477, 763

\bibitem[{{Schechter}(1976)}]{1976ApJ...203..297S}
{Schechter}, P. 1976, \apj, 203, 297

\bibitem[{{Schmitt}(2001)}]{2001AJ....122.2243S}
{Schmitt}, H.~R. 2001, \aj, 122, 2243

\bibitem[{{Scoville et al.}(2007)}]{2007ApJS..172....1S}
{Scoville et al.}, N. 2007, \apjs, 172, 1

\bibitem[{{Sharp}(2006)}]{2006SPIE.6269E..14S}
{Sharp}, R. 2006, in Presented at the Society of Photo-Optical Instrumentation
  Engineers (SPIE) Conference, Vol. 6269, Society of Photo-Optical
  Instrumentation Engineers (SPIE) Conference Series

\bibitem[{{Stott} {et~al.}(2007){Stott}, {Smail}, {Edge}, {Ebeling}, {Smith},
  {Kneib}, \& {Pimbblet}}]{2007ApJ...661...95S}
{Stott}, J.~P., {Smail}, I., {Edge}, A.~C., {Ebeling}, H., {Smith}, G.~P.,
  {Kneib}, J., \& {Pimbblet}, K.~A. 2007, \apj, 661, 95

\bibitem[{{Strateva et al.}(2001)}]{2001AJ....122.1861S}
{Strateva et al.}, I. 2001, \aj, 122, 1861

\bibitem[{{Tanaka} {et~al.}(2005){Tanaka}, {Kodama}, {Arimoto}, {Okamura},
  {Umetsu}, {Shimasaku}, {Tanaka}, \& {Yamada}}]{2005MNRAS.362..268T}
{Tanaka}, M., {Kodama}, T., {Arimoto}, N., {Okamura}, S., {Umetsu}, K.,
  {Shimasaku}, K., {Tanaka}, I., \& {Yamada}, T. 2005, \mnras, 362, 268

\bibitem[{{Tanaka} {et~al.}(2009){Tanaka}, {Lidman}, {Bower}, {Demarco},
  {Finoguenov}, {Kodama}, {Nakata}, \& {Rosati}}]{2009A&A...507..671T}
{Tanaka}, M., {Lidman}, C., {Bower}, R.~G., {Demarco}, R., {Finoguenov}, A.,
  {Kodama}, T., {Nakata}, F., \& {Rosati}, P. 2009, \aap, 507, 671

\bibitem[{{van Dokkum} \& {Franx}(2001)}]{2001ApJ...553...90V}
{van Dokkum}, P.~G. \& {Franx}, M. 2001, \apj, 553, 90

\bibitem[{{Weiner et al.}(2005)}]{2005ApJ...620..595W}
{Weiner et al.}, B.~J. 2005, \apj, 620, 595

\bibitem[{{Weinmann} {et~al.}(2006){Weinmann}, {van den Bosch}, {Yang}, \&
  {Mo}}]{2006MNRAS.366....2W}
{Weinmann}, S.~M., {van den Bosch}, F.~C., {Yang}, X., \& {Mo}, H.~J. 2006,
  \mnras, 366, 2

\bibitem[{{Willmer et al.}(2006)}]{2006ApJ...647..853W}
{Willmer et al.}, C.~N.~A. 2006, \apj, 647, 853

\bibitem[{{Wolf} {et~al.}(2005){Wolf}, {Gray}, \&
  {Meisenheimer}}]{2005A&A...443..435W}
{Wolf}, C., {Gray}, M.~E., \& {Meisenheimer}, K. 2005, \aap, 443, 435

\bibitem[{{Wolf} {et~al.}(2003){Wolf}, {Meisenheimer}, {Rix}, {Borch}, {Dye},
  \& {Kleinheinrich}}]{2003A&A...401...73W}
{Wolf}, C., {Meisenheimer}, K., {Rix}, H., {Borch}, A., {Dye}, S., \&
  {Kleinheinrich}, M. 2003, \aap, 401, 73

\bibitem[{{Wolf et al.}(2009)}]{2009MNRAS.393.1302W}
{Wolf et al.}, C. 2009, \mnras, 393, 1302

\bibitem[{{Wyder et al.}(2007)}]{2007ApJS..173..293W}
{Wyder et al.}, T.~K. 2007, \apjs, 173, 293

\bibitem[{{Yee}(1991)}]{1991PASP..103..396Y}
{Yee}, H.~K.~C. 1991, \pasp, 103, 396

\bibitem[{{Yee} {et~al.}(2007){Yee}, {Gladders}, {Gilbank}, {Majumdar},
  {Hoekstra}, \& {Ellingson}}]{2007ASPC..379..103Y}
{Yee}, H.~K.~C., {Gladders}, M.~D., {Gilbank}, D.~G., {Majumdar}, S.,
  {Hoekstra}, H., \& {Ellingson}, E. 2007, in Astronomical Society of the
  Pacific Conference Series, Vol. 379, Cosmic Frontiers, ed. {N.~Metcalfe \&
  T.~Shanks}, 103--+

\bibitem[{{Yee} \& {Green}(1987)}]{1987ApJ...319...28Y}
{Yee}, H.~K.~C. \& {Green}, R.~F. 1987, \apj, 319, 28

\bibitem[{{Yee} {et~al.}(2005){Yee}, {Hsieh}, {Lin}, \&
  {Gladders}}]{2005ApJ...629L..77Y}
{Yee}, H.~K.~C., {Hsieh}, B.~C., {Lin}, H., \& {Gladders}, M.~D. 2005, \apjl,
  629, L77

\end{thebibliography}
\end{document}